\documentclass[fleqn,usenatbib]{mnras}

\usepackage[T1]{fontenc}
\usepackage{ae,aecompl}

\usepackage{graphicx}	
\usepackage{amsmath}	
\usepackage{amssymb}	

\def\lya{{\rm\,Ly$\alpha$}}
\def\ha{{\rm\,H$\alpha$}}
\def\hb{{\rm\,H$\beta$}}

\def\oii{{\rm\,[O{\sc ii}]}}

\def\oiii{{\rm\,[O{\sc iii}]}}
\def\civ{{\rm\,C{\sc iv}}}
\def\ciii{{\rm\,C{\sc iii}]}}
\def\mgii{{\rm\,Mg{\sc ii}}}

\def\hi{{\rm\,H{\sc i}}}

\def\msun{{\rm M}$_{\odot}$}

\title[Similarities and uniqueness of Ly$\alpha$ emitters]
{Similarities and uniqueness of Ly$\boldsymbol\alpha$ emitters among star-forming galaxies at $\boldsymbol{z}$=2.5}

\author[R. Shimakawa et al.]{
Rhythm Shimakawa,$^{1,2}$\thanks{rhythm.shimakawa@nao.ac.jp} 
Tadayuki Kodama,$^{1,2}$\thanks{t.kodama@nao.ac.jp}
Takatoshi Shibuya,$^{3}$
\newauthor
Nobunari Kashikawa,$^{1,2}$
Ichi Tanaka,$^{4}$
Yuichi Matsuda,$^{1,2}$
Ken-ichi Tadaki,$^{5}$
\newauthor
Yusei Koyama,$^{1,4}$
Masao Hayashi,$^{2}$
Tomoko L. Suzuki$^{1,2}$
and Moegi Yamamoto$^{1}$
\\
$^{1}$Department of Astronomical Science, SOKENDAI, Osawa, Mitaka, Tokyo 181-8588, Japan\\
$^{2}$National Astronomical Observatory of Japan, Osawa, Mitaka, Tokyo 181-8588, Japan\\
$^{3}$Institute for Cosmic Ray Research, The University of Tokyo, 5-1-5 Kashiwanoha, Kashiwa, Chiba 277-8582, Japan\\
$^{4}$Subaru Telescope, National Astronomical Observatory of Japan, 650 North A'ohoku Place, Hilo, HI 96720, USA\\
$^{5}$Max-Planck-Institut f\"{u}r Extraterrestrische Physik, Giessenbachstrasse, D-85748 Garching Germany
}

\date{Accepted 2017 January 7. Submitted 2015 December 18; in original form 2017 January 7}

\pubyear{2017}

\begin{document}
\label{firstpage}
\pagerange{\pageref{firstpage}--\pageref{lastpage}}
\maketitle

\begin{abstract}
We conducted a deep narrow-band imaging survey with the Subaru Prime 
Focus Camera on the Subaru Telescope and constructed a sample of 
Ly$\alpha$ emitters (LAEs) at $z=2.53$ in the UDS-CANDELS field where 
a sample of H$\alpha$ emitters (HAEs) at the same redshift is already 
obtained from our previous narrow-band observation at NIR. The deep 
narrow-band and multi broadband data allow us to find LAEs of stellar
masses and star-formation rates (SFRs) down to $\gtrsim$$10^8$
M$_\odot$ and $\gtrsim$0.2 M$_\odot$/yr, respectively. We show that 
the LAEs are located along the same mass-SFR sequence traced by normal 
star-forming galaxies such as HAEs, but towards a significantly lower 
mass regime. Likewise, LAEs seem to share the same mass--size relation 
with typical star-forming galaxies, except for the massive LAEs, which 
tend to show significantly compact sizes. We identify a vigorous mass 
growth in the central part of LAEs: the stellar mass density in the 
central region of LAEs increases as their total galaxy mass grows. 
On the other hand, we see no Ly$\alpha$ line in emission for most of 
the HAEs. Rather, we find that the Ly$\alpha$ feature is either absent 
or in absorption (Ly$\alpha$ absorbers; LAAs), and its absorption 
strength may increase with reddening of the UV continuum slope. We 
demonstrate that a deep Ly$\alpha$ narrow-band imaging like this 
study is able to search for not only LAEs but also LAAs in a certain 
redshift slice. This work suggests that LAEs trace normal 
star-forming galaxies in the low-mass regime, while they remain as a 
unique population because the majority of HAEs are not LAEs.
\end{abstract}




\section{Introduction}

The redshift interval of $z$=2.1--2.6 is the key epoch for us to understand 
the formation mechanisms of star-forming galaxies. In this era, the cosmic 
star-formation density is peaked \citep{Hopkins:2006}, galaxy morphology is 
still under construction \citep{Papovich:2005}, and gaseous flows (feeding 
and feedback) are more prevalent compared to lower redshift counterparts 
\citep{Yabe:2015}. Galaxies at this epoch allow us to study the physical 
origins of these processes based on their strong emission lines such as 
\ha$\lambda6563$, \oiii$\lambda5007$, \oii$\lambda3727$, and \lya$\lambda1216$ 
with the optical and near-infrared imagers and spectrographs. For example, 
we can directly investigate the physical properties of \lya\ emitters (LAEs) 
in this redshift range such as gaseous metallicities and ionization 
parameters \citep{Nakajima:2013}. We can also compare the characteristics 
of LAEs with those of \ha\ emitters (HAEs) \citep{Hayes:2010,Oteo:2015} up 
to $z\sim2.5$ which are relatively unbiased sample of star-forming galaxies. 
At higher redshifts, a comparison can be made with Lyman break galaxies 
(LBGs; \citealt{Verhamme:2008,Malhotra:2012}). Such comparison analyses 
should unveil the physical characteristics of LAEs. In addition, the 
observations of nearby LAE analogues also help us resolve the physical 
mechanism of \lya\ radiative transfer in detail 
\citep{Kunth:1998,Atek:2009b,Hayes:2013,Hayes:2015,Rivera:2015}. 

The properties of LAEs provide information on the reionisation history in 
the early Universe (e.g. \citealt{Dawson:2007,Kashikawa:2006,Stark:2010}), 
although the intrinsic properties of LAEs may also change with redshift 
\citep{Nilsson:2011}. The past analyses find that LAEs tend to be younger, 
less massive, and less dusty star-forming galaxies (e.g. 
\citealt{Gawiser:2006,Nilsson:2007,Ouchi:2008,Guaita:2011}). The escape 
fraction of \lya\ photons and the ionization parameter are both high in these 
systems, compared to other star-forming galaxies 
\citep{Nakajima:2012,Nakajima:2013,Oteo:2015}. Note that a minor fraction of
LAEs can constitute massive and red galaxies (see also \citealt{Oteo:2012, 
Bridge:2013,Sandberg:2015,Finkelstein:2015,Taniguchi:2015}). Furthermore, 
LAEs tend to lie above the tight star-formation rate (SFR) vs. stellar mass 
relation called the main sequence of star-forming galaxies 
(\citealt{Vargas:2014, Hagen:2014}, but see also 
\citealt{Nilsson:2011b,Song:2014,Kusakabe:2015,Finkelstein:2015}).
LAEs tend to have more compact star-forming regions seen in the rest-frame UV 
light compared to LBGs, and their angular sizes of UV continuum are 
independent of redshift at $z<6$ \citep{Malhotra:2012}. A more systematic 
study has been carried out by \citet{Hathi:2016}, who compare the 
physical properties among non-LAEs, LAEs with small equivalent width 
(EW$<20$ \AA), and those with EW$>20$ \AA. They have shown that the UV slope 
and the UV magnitude at 1500 \AA, M$_{1500}$, of star-forming galaxies
with and without \lya\ emission lines are similar for a given i-band 
magnitude limit ($<25$ mag). 
In addition, \citet{Hagen:2016} have found no statistical difference 
in physical and morphological parameters such as specific SFR, SFR surface
density, half-light radius and \oiii\ equivalent width between LAEs and 
non-LAEs at $z\sim2$, based on the sample limited to the rest-frame optical
line flux of $\gtrsim10^{-17}$ erg/s/cm$^2$.

However, a careful consideration is needed to discuss representative
characteristics of LAEs. The excess of SFRs in LAEs with respect to the 
main sequence can be interpreted in such a way that a large amount of 
ionizing photons from young and active starbursts may allow \lya\ photons 
to escape from the systems more easily.  On the other hand, various 
previous studies suggest that the properties of LAEs are quite diverse, 
which may imply that the escape of \lya\ photons is a stochastic process 
rather than an ordered duty cycle (see also \citealt{Nagamine:2010}). 
Moreover, the escape of \lya\ photons depends not only on the energy input 
from star-forming regions or active galactic nuclei but also on the amount 
of dust and circumgalactic medium 
\citep{Verhamme:2012,Yajima:2012,Shibuya:2014a,Shibuya:2014b}.
Larger gas covering fraction leads to higher line of sight extinctions. 
\citet{Reddy:2016b} have found that those quantities correlates with UV 
continuum slope in the sense of redder UV continua tend to have higher 
covering fraction of \hi\ and dust. In addition, the effect of stellar 
absorption is also a non-negligible factor \citep{Pena:2013}. Another 
concern is that all the previous LAE studies have been limited by the 
\lya\ limiting flux that can be reached in the observations. The \lya\ 
luminosity depends on galaxy populations, and in fact most of the very 
bright LAEs ($\gtrsim3\times10^{43}$ erg/s) host active galactic nuclei 
\citep{Ouchi:2008,Konno:2016,Sobral:2016}.
For example, we may find LAEs on the main sequence of 
normal star-forming galaxies when we conduct deeper observations.
If most of the LAEs have normal specific SFRs similar to those of
typical star-forming galaxies, the escape of the \lya\ photons cannot
be due simply to their high star-formation activity, and some other
factors would be more dominant.

To understand the nature of LAEs, the dual emitter survey of HAEs and 
LAEs at the same redshift is very powerful. \citet{Hayes:2010} first 
reported that only six out of 55 \ha-selected galaxies were detected 
by a narrowband filter for \lya\ line at $z=2.2$. They also found 
very low \lya\ photon escape fractions of $\sim5\%$, which is 
consistent with another recent dual survey of HAEs and LAEs 
\citep{Matthee:2016,Sobral:2016}. Those studies noted that it is 
difficult to account for such a low \lya\ escape fraction only by dust 
attenuation (see also \lya\ line observations of nearby galaxies e.g. 
\citealt{Atek:2008}). All these studies point towards the fact that 
most of the star-forming galaxies do not actually show \lya\ emission 
lines that are strong enough to be detected by a standard narrowband 
imaging observation. What if we go significantly deeper in \lya\ 
narrowband imaging observation? The fraction of dual emitters (HAEs 
and LAEs) should depend on the survey depths. Indeed, the most of the 
past \lya\ line imaging surveys with NB$_\mathrm{5\sigma}\lesssim25$ 
mag do not reach continua at \lya\ line wavelength for many 
\ha-selected galaxies at $z>2$, in particular for dusty galaxies. 
Deeper \lya\ observation will give us more comprehensive insights 
into the \lya\ emissivities of high redshift galaxies.

Our project named MAHALO-Subaru (Mapping HAlpha Line of Oxygen with 
Subaru) has successfully mapped out \ha, \oii, and \oiii\ emitters at 
$z$=0.4--3.8 in 14 overdense regions (e.g. \citealt{Kodama:2004, 
Koyama:2013a,Hayashi:2014}) and some blank fields in COSMOS- and UDS- 
CANDELS fields (e.g. \citealt{Tadaki:2013, Suzuki:2015}). Some 
follow-up spectroscopic surveys have been conducted by 
\citet{Shimakawa:2014,Shimakawa:2015,Shimakawa:2015b}. We have now 
recently conducted a deep narrow-band survey of \lya\ emitters with 
Suprime-Cam in coordination with the existing companion survey of \ha\ 
emitters with MOIRCS \citep{Tadaki:2013}. This enables us to probe 
star-forming galaxies over a wider mass range and across various 
environments at $2<z<2.6$, where both \ha\ and \lya\ lines can be 
accessed with the ground-based telescopes.

This {\it Paper} studies the LAEs at $z=2.53$ in the UDS-CANDELS field 
based on the deep NB imaging observations with Suprime-Cam, as the first 
such targets. We mainly focus on star-forming activities and stellar 
sizes of the LAEs, as compared to HAEs. The details of the observations 
and the analyses can be found in \S2. \S3 presents the star-forming 
activities and stellar sizes of LAEs, and compares them with those of 
HAEs, which are more representative, massive star-forming galaxies at the 
same redshift. Discussion is given in \S4, and we conclude this work in 
\S5. We assume the cosmological parameters of $\Omega_M$=0.3, 
$\Omega_\Lambda$=0.7 and $H_0$=70 km/s/Mpc and employ a 
\citet{Chabrier:2003} stellar initial mass function (IMF). The AB 
magnitude system \citep{Oke:1983} is used throughout this paper.

\newpage

\section{Observations and data analyses}

\begin{figure}
	\centering
	\includegraphics[width=75mm]{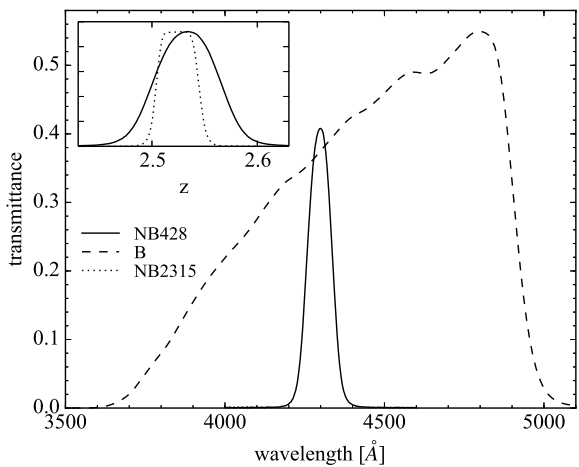}
    \caption{The black solid and the dashed curves indicate the filter 
    response curves of NB428 and B-band on the Suprime-Cam, respectively. 
    The upper left inset shows the filter response curves of NB428 on 
    Suprime-Cam (solid curve) and NB2315 on MOIRCS (dotted curve) as a 
    function of redshift for \lya\ and \ha\ lines, respectively.}
    \label{fig1}
\end{figure}

The narrow-band (NB) filter, NB428, is installed on the Subaru Prime 
Focus Camera (Suprime-Cam; \citealt{Miyazaki:2002}). It is a user filter 
belonging to R. Shimakawa (P.I. of this {\it Paper}) and is manufactured 
specifically for the purpose of searching for \lya\ emission and 
absorption lines at $z$=2.53$\pm$0.03. Its central wavelength and FWHM 
are 4297 \AA\ and 84\AA, respectively. It has a good spatial homogeneity 
of the central wavelength (less than $\pm3$ \AA) across the field of 
view. The NB428 filter makes a pair with our companion filter, NB2315, 
(belonging to Tadayuki Kodama; see \citealt{Hayashi:2012, Tadaki:2013}) 
which is installed on MOIRCS/Subaru. The NB2315 filter can sample \ha\ 
emitting star-forming galaxies at the same redshift of LAEs down to a 
\ha\ flux limit of 2.6$\times$10$^{-17}$ erg/s/cm$^2$. The filter 
response curves are shown in Figure \ref{fig1}.

LAEs at $z$=2.53 selected with the NB428 filter also have some great 
advantages. Those broad-band photometries of H$_\mathrm{F160W}$ and 
K$_s$ do not include \hb$\lambda4861$, \oiii$\lambda\lambda4959,5007$ or 
\ha$\lambda6563$ lines. This allows us to measure their accurate stellar 
masses and sizes from these broad-band data. \lya$\lambda1216$ line is 
also out of the response curve of the U-band of MegaCam on the Canada 
France Hawaii Telescope (CFHT). Moreover, all strong optical emission 
lines such as \ha, \hb, \oiii, and \oii\ lines including the 
\oiii$\lambda4363$ auroral line can be observed from the ground-based 
telescopes, which enable us to derive the robust measurements of gaseous 
metallicities based on the direct method \citep{Izotov:2006} by 
spectroscopic observations with future larger aperture telescopes.

\subsection{Narrow-band imaging and data reduction}

\begin{figure}
	\includegraphics[width=\columnwidth]{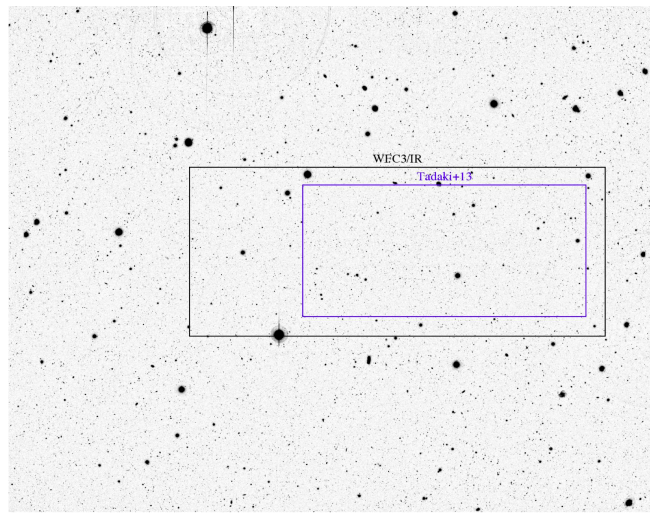}
    \caption{The reduced, full NB428 image (34'$\times$27').
    This work concentrates on the overlapped area with the HST F160W 
    image (WFC3/IR) which is marked with the black square 
    (22'$\times$9'). We also incorporate the sample of \ha\ emitters 
    at $z$=2.5 constructed within the narrower field shown by the 
    blue square.}
    \label{fig2}
\end{figure}

We performed a NB428 imaging observation with Suprime-Cam on UDS 
(Ultra Deep Survey; \citealt{Lawrence:2007,Foucaud:2007}) field on 
2014 August 19, as a Subaru Service Program (S14B-198S; R. Shimakawa 
et al.). The wide field of view of Suprime-Cam (FoV=34'$\times$27') 
covers the entire area of the UDS-CANDELS-WFC3/IR field 
(\citealt{Grogin:2011,Koekemoer:2011}; Fig.~\ref{fig2}), where 
archive data are available including deep, high-resolution HST images. 
The NB observation was carried out in a standard manner with an 
individual exposure time of 1200 sec and two sets of five dithering 
points on a 1 arcmin radius circle. The total integration time 
amounted to 3.3 hrs, and the images were taken under a photometric 
sky condition with the seeing sizes of 0.62--0.86 arcsec, and the 
typical airmass of 1.2. 

The obtained images were reduced by using a data reduction package 
for the Suprime-Cam, {\sc sdfred} (ver.2; 
\citealt{Yagi:2002,Ouchi:2004}). This produces a well-reduced final 
image semi-automatically. The actual procedures consist of bias 
subtractions, flat fielding, distortion correction, PSF (point spread 
function) matching, sky subtraction, and image combining. More 
details of the pipeline are described in \citet{Ouchi:2004}. We also 
implement cosmic ray reduction using the algorithm of cosmic ray 
identification, L.A.Cosmic \citep{Dokkum:2001}. The final combined 
image has a seeing size of FWHM=0.85 arcsec, and the $5\sigma$ 
limiting magnitude of 26.63 in 1.5 arcsec diameter aperture. The zero 
point magnitude of 31.490 in 1200 sec is determined from a 
spectroscopic standard star, Feige 110, whose high-resolution 
spectrum was obtained by X-shooter on the Very Large Telescope 
\citep{Moehler:2014}. 

\subsection{Colour magnitude diagram}

\begin{figure}
	\centering
	\includegraphics[width=75mm]{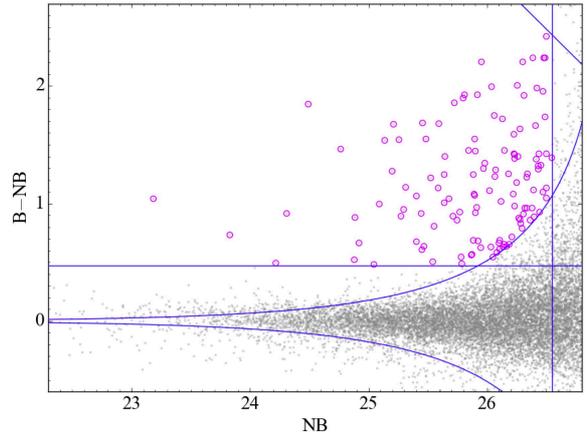}
    \caption{The colour-magnitude diagram of NB428 vs. B$-$NB428. 
    The grey dots represent all the NB-detected galaxies in the UDS-WFC3
    field. Stars with F160W$<$25 mag classified by \citet{Skelton:2014} 
    have been removed. The blue curves show the threshold of $\pm3\sigma$ 
    flux excess. The blue vertical and horizontal lines indicate the 
    $5\sigma$ limiting magnitude and the criterion used for the observed 
    equivalent width (EW$_\mathrm{obs}$=53\AA), respectively. The blue 
    diagonal line indicates 2$\sigma$ limit of B-band. The purple open 
    circles show the NB emitters that satisfy our selection criteria.} 
    \label{fig3}
\end{figure}

We first identify narrow-band emitters by searching for objects that 
show flux excesses in the narrow-band (NB428) as compared to the 
broad-band (B). PSF size is matched to the NB428 image (FWHM=0.85 
arcsec) and pixel scale is adjusted to mosaic image of B-band (0.268 
arcsec/pixel) distributed by \citet{Cirasuolo:2010}. Hereafter we only 
use the data covering the UDS-CANDELS-WFC3/IR field (Fig.~\ref{fig2}). 
The source identification is made using the astronomical software for 
source extraction called SExtractor \citep{Bertin:1996}. Source 
detections are performed with the double-image mode of the SExtractor 
with detection parameters of {\sc detect\_minarea}=4, 
{\sc detect\_thresh}= 1.5, {\sc analysis\_thresh}= 2. We use a 1.5 
arcsec diameter aperture both in NB428 and B-band to measure the 
object colours. Those 5$\sigma$ limiting magnitudes correspond to 
NB$_\mathrm{5\sigma}$=26.55 and B$_\mathrm{5\sigma}$=27.89 including 
the galactic extinction. The galactic extinctions of 
A$_\mathrm{B}$=0.081 and A$_\mathrm{NB428}$=0.082 mag are assumed 
based on the \citet{Fitzpatrick:1999} extinction law with 
R$_\mathrm{V}$=3.1 and E(B$-$V) from \citet{Schlafly:2011}, which 
is tailored to the specification in the 3D-HST catalogue 
\citep{Skelton:2014} that we use (\S2.3). NB emitters are then 
selected by using a colour magnitude diagram of NB428 vs. B$-$NB 
(Fig.~\ref{fig3}). We here apply a small correction to the NB 
magnitudes by $-0.1$ mag to remove an offset seen in the median 
B$-$NB colour of the bright galaxies (NB=24--25 mag). This offset 
would be mostly caused by a systematic zero-point offset 
due to the colour terms of individual galaxies. The 
amount of the colour terms are actually consistent with those 
predicted by the galaxies at the photometric redshift around 2.5 
(see Appendix B). Because of this, this work does not consider a 
negligible colour term that originates from the small offset of the 
central wavelengths (170 \AA) between NB428 and B-band. It actually 
has a strong benefit for searching LAEs having faint UV luminosities 
compared to similar studies for LAEs at slightly lower-$z$ which have 
used shallower U-band data (U$_\mathrm{5\sigma}$=26.63 in the case of 
the UDS-CANDELS field). Then, we set the selection criteria: the 
observed equivalent width (EW$_\mathrm{obs}$) is greater than 53 \AA\ 
corresponding to 15 \AA\ in the rest frame at $z=2.53$, B$-$NB$>$0.476 
mag or $\sim4\sigma$ scatter of NB$-$B for the luminous galaxies, the 
limiting magnitudes of $5\sigma$ ($2\sigma$) in NB (B) band, and the 
NB flux shows an excess of more than 3$\sigma$ compared to the B-band 
at 1.5 diameter aperture photometry, which is equivalent to a NB flux 
of 8.52$\times$10$^{-18}$ erg/s/cm$^2$ and a NB luminosity of 
4.40$\times$10$^{41}$ erg/s if they are located at $z=2.53$. We should 
note that B-band magnitude cut is not critical to our results and 
conclusions, since only two objects additionally selected as NB 
emitters without this threshold cannot pass all the selection criteria 
in this paper. In total, 123 galaxies satisfied these criteria, and 
they are represented by magenta circles in Fig.~\ref{fig3}. 

\subsection{Catalogue matching}

Since one motivation in this paper is to study physical 
properties of LAEs at $z=2.5$, multi-band photometry is required 
to derive their stellar masses, SFRs and so on. In particular, 
deep H$_\mathrm{F160W}$-band photometry from the CANDELS 
\citep{Grogin:2011,Koekemoer:2011} is critical to estimate robust 
stellar masses of low-mass LAEs at $z$=2.5. 
We perform a cross-matching of our NB-selected emitters to the 
sources in the 3D-HST catalogue \citep{Skelton:2014} which is 
constructed from the WFC3/HST images (F125W/F140W/F160W) taken by 
the CANDELS project \citep{Grogin:2011, Koekemoer:2011} and the 
3D-HST Treasury Survey \citep{Brammer:2012,Skelton:2014}. We 
employed a variety of photometric data in this catalogue covering 
a wide wavelength range from various imaging data; CFHT U-band 
(Almaini/Foucaud et al. in preparation), Subaru B,V,R,i,z-band 
\citep{Furusawa:2008}, HST F606 and 814W \citep{Koekemoer:2011}, 
UKIRT J,H,Ks (\citealt{Lawrence:2007}; Almaini et al. in 
preparation), HST F140W \citep{Brammer:2012}, F125 and 160W 
\citep{Koekemoer:2011}. The catalogue includes the total fluxes 
correcting for the flux loss and also photometric zero-points in 
an empirical manner (see \citealt{Skelton:2014} for details). We 
first adjust the WCS information of the NB image to the combined 
image of WFC3/IR bands (F125W/F140W/F160W) made by 3D-HST based 
on the {\sc iraf}\footnotemark[1] task; {\sc ccmap} and 
{\sc ccsetwcs}. The coordinated WCS data of the NB image is 
well matched to those of WFC3 within less than 0.1 arcsec offset 
according to the positions of bright stars in each image. We 
then cross-matched the NB-selected sources to the 3D-HST catalogue, 
and select the galaxy closest to each NB emitter within a 0.4 
arcsec radius. For 69 sources out of 123 NB emitters, we can find 
the counterparts. Most of the removed 54 objects should be very 
low-mass objects that are too faint even in the WFC3 image. Indeed, 
most of the catalogue-matched NB emitters are detected only at 
F125W and F160W bands in the NIR regime with more than a few sigma 
significance in 0.7 arcsec aperture diameter photometry. This work 
subsequently removes 11 very faint objects at F160W$>26.89$ that is 
2$\sigma$ ($\sim10\sigma$) limiting magnitude in 0.7 (0.35) arcsec 
aperture photometry (according to 
\citealt{Skelton:2014, Shibuya:2014a}). This value roughly 
corresponds to the stellar mass of $1\times10^{8}$ solar mass at 
$z=2.53$.

\footnotetext[1]{{\sc iraf} is distributed by National 
Optical Astronomy Observatory and available at \url{iraf.noao.edu}}

Attentive considerations could be needed since spatial offsets of 
\lya\ peak (by a few arcsec at $z=2.2$) from the stellar continuum 
are often seen as shown by \citet{Shibuya:2014a}. Indeed, the peak 
locations of NB fluxes of some selected NB emitters seem to deviate 
from those of stellar continua seen in the WFC3 images. However, 
most of the NB emitters are as faint as $\sim$ 26 mag (less than 
10$\sigma$ detections) in the NB magnitude. Thus it is hard to 
identify the spatial peaks in NB fluxes of those faint galaxies 
based only on the seeing-limited data. Furthermore, we also find 
that the separation angles between the peaks of NB fluxes and 
stellar continua of the NB emitters moderately correlate with their 
angular sizes, which means that flux peak locations of diffuse 
objects are not accurately measured. Because of this problem, we 
cannot resolve the spatial peaks of \lya\ emission robustly. This 
work employs the NB emitters, which were cross-matched to WFC3 
source within 0.4 arcsec angular distance. This threshold is 
determined by the fact that 95\% of the catalogue-matched galaxies 
are located within this criterion when we conduct the cross-matching 
by match criteria within 1.0 arcsec separation between NB sources and 
the 3D-HST sources. 

Those processes were conducted using the Tool for OPerations on 
Catalogues And Tables ({\sc topcat}; \citealt{Taylor:2005, Taylor:2015}).

\subsection{Sample selection}

\begin{figure}
	\centering
	\includegraphics[width=75mm]{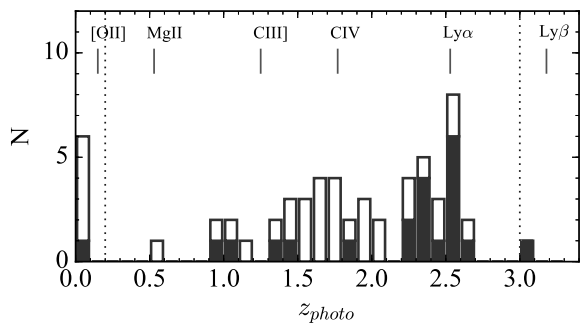}
    \caption{Photometric redshift distribution of the NB emitters by 
    \citet{Momcheva:2016}. We employ the photometric redshift 
    $z_\mathrm{best}$ in their catalogue. Filled and opened histograms
    present the NB emitters at F160W$\le$25 and F160W$>$25 mag, 
    respectively. This work simply remove the objects at photometric 
    redshifts of $z_\mathrm{best}\le0.2$ or $3.0\le z_\mathrm{best}$.}
    \label{fig4}
\end{figure}

While the majority of them are likely to be LAEs at $z$=2.53$\pm$0.03, 
we need to remove some candidates, especially foreground \oii\ line 
emitters at $z$=0.15 and background Lyman break galaxies and Ly$\beta$ 
emitters at $z>3$. We employed a photometric redshift catalogue by 
\citet{Momcheva:2016}, which is based on the combined technique of 
multi-band photometries and grism data by the 3D-HST Treasury Program.  
We simply excluded seven candidates of such contaminations showing 
photo-$z$ ($z_\mathrm{best}$) of $z_\mathrm{best}\le0.2$ or 
$3.0\le z_\mathrm{best}$ in their catalogue. One should note that the 
redshift estimation by \citet{Momcheva:2016} does not work well for 
our NB emitters because of their faintness at NIR bands, and thus 
photo-$z$ estimation should have larger uncertainties relative to 
their entire 3D-HST sample ($\sigma_z\sim0.003\times(1+z)$ in $JH<24$). 
Actually photo-$z$ distribution of the NB emitters at F160W$>$25 mag 
systematically shifted toward $z<2$ as seen in the Figure~\ref{fig4}. 
However, we believe that some of these would be LAEs at $z=2.5$ rather 
than foreground line emitters, because this trend has been reported by 
\citet{Shivaei:2015} which show that estimation of photometric 
redshifts are systematically underestimated for faint at F160W and/or 
blue galaxies at $z$=2--2.6 like our LAE sample. On another front, 
it is highly likely that such \mgii$\lambda2800$, \ciii$\lambda1909$, 
and \civ$\lambda1550$ emitters (at $z$=0.53, 1.25, and 1.77, 
respectively) are contained in LAE candidates especially for those at
the bright end of the \lya\ luminosity function as noted by 
\citet{Sobral:2016}. Because of the lack of spec-$z$ confirmation of our
LAE samples, we ignore these possible contaminations. However, we stress 
that our conclusions remain consistent even if we select the LAEs only 
within the photometric redshifts range of $z$=2.3--2.7. Moreover, this 
work mainly focuses on the physical properties of low-mass LAEs, and thus  
the contamination of bright foreground emitters may not be a big issue.

Finally, we carefully inspected multi band photometries of LAE 
candidates in order to check whether unrealistic objects still blend 
into our sample. We then find an object showing strange i-band excess 
($\gtrsim4\sigma$) relative to bluer and redder band photometries. It 
is highly likely that foreground \oii\ emitters with $\gtrsim50$ \AA\ 
also have high EW of \ha\ line ($\lambda_\mathrm{obs}\sim7560$ \AA), 
which should cause i-band excess. Thus, this object considered a 
foreground \oii\ emitter is also removed from the sample. 
Eventually, we obtained 50 LAE candidates at $z=2.5$. These selection 
processes are summarized in table~\ref{tab1}. 

\begin{table}
	\caption{Selection proceeding of LAE candidates at $z=2.53$.}
 	\label{tab1}
	\begin{tabular}{lrr}
		\hline
		Selection criteria & removed & remained \\
		\hline
        NB emitters & --- &  123 \\ 
        Catalogue matching with 3D-HST ($<$0.4") & 54 &  69 \\ 
        F160W/WFC3 magnitude cut ($\le$26.89) & 11 & 58 \\ 
        Photometric redshift cut ($0.2<z<3.0$) & 7 & 51 \\ 
        i-band excess (contaminant \oii\ emitters) & 1 & 50 \\ 
        \hline
    \end{tabular}
\end{table}

\subsection{Final catalogue}

The sample employed in this work consists of 50 LAEs at $z$=2.5. 
These are limited by the \lya\ luminosity ($>$4.40$\times$10$^{41}$ 
erg/s), EW ($>$15 \AA), and H$_\mathrm{F160W}$ band magnitude 
($<$26.89 mag). Hereafter, we basically use their photometries from 
the 3D-HST catalogue \citep{Skelton:2014} except their \lya\ 
luminosities and equivalent widths (EWs), unless otherwise noted. 

This work also collects HAE samples at the same redshift taken by 
the past narrow-band imaging survey of the UDS-CANDELS field with 
the MOIRCS on the Subaru telescope \citep{Tadaki:2013, Tadaki:2014}. 
The data is limited to their observed \ha\ flux above 
$2.6\times10^{-17}$ erg/s/cm$^2$, and more than EW$_\mathrm{H\alpha}$ 
of 40 \AA\ in \ha\ emission at $z$=2.53. About half of them are 
spectroscopically confirmed by MOIRCS, KMOS, and ALMA. More details 
are described in \citet{Tadaki:2013,Tadaki:2014,Tadaki:2015}, and 
forthcoming papers. We employ 37 out of 44 HAEs with B-band 
detections more than 2 sigma (28.88 mag) with 1.5 arcsec diameter 
photometry by the SExtractor. Two of those are also selected as LAEs 
by this observation. The B-band magnitude cut actually allows us to 
miss seven heavily obscured HAEs. However, this work neglects them in 
order to be consistent with the sample selection of LAEs, and also 
since we cannot constrain EWs of \lya\ of such dusty HAEs at all. 
After that, the HAE sample were cross-matched to the 3D-HST catalogue 
in the same manner as LAEs. Therefore, multi-band photometries of 
HAEs used in this work are a bit different from those reported by 
past studies.

\begin{figure}
	\centering
	\includegraphics[width=75mm]{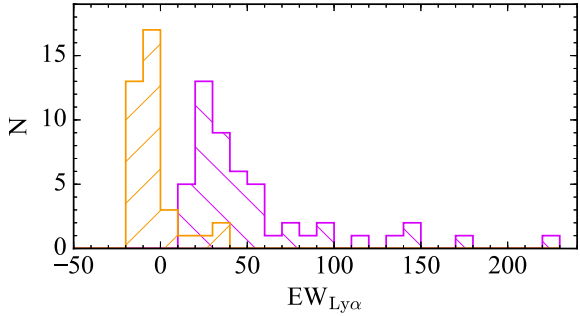}
    \caption{EW$_\mathrm{Ly\alpha}$ distribution of LAEs (purple 
    backslash) and HAEs (orange forward slash). Here, positive 
    values show emissions and negative values mean absorptions.}
    \label{fig5}
\end{figure}

Figure~\ref{fig5} represents EW$_\mathrm{Ly\alpha}$ distributions
of LAE and HAE samples. For HAEs without NB428 detections, we just 
insert 2 sigma upper limits. The stellar population model by 
\citet{Malhotra:2002} based on the Starburst99 \citep{Leitherer:1999} 
cannot explain extremely high EW$_\mathrm{Ly\alpha}>240$ \AA\ of 
\lya\ emission without the assumption of extreme IMF (much heavier 
IMF) (see also 
\citealt{Charlot:1993,Schaerer:2002,Schaerer:2003,Laursen:2013}). 
Fig.~\ref{fig5} indicates that our LAE selection does not contain 
such objects. On the other hand, we could not find strong \lya\ 
emission from the most of HAEs. Rather, we see a deficit in NB428 
magnitude compared to B-band for a part of our HAE samples, which 
means that \lya\ line is in absorption rather than in emission for 
them. EW$_\mathrm{Ly\alpha}$ distribution of HAEs seems consistent 
with that of Lyman-break galaxies at the similar redshift within 
the errors \citep{Reddy:2008}. The details of such \lya\ absorbers 
and their possible origins are described in \S3.3 and \S4.3. 

We then conducted SED-fitting based on the SED-fitting code 
{\sc fast} distributed by \citet{Kriek:2009}. The SED fitting was 
carried out assuming a fixed redshift of $z$=2.53, the stellar 
population model of \citet{Bruzual:2003}, the \citet{Calzetti:2000} 
attenuation curve, the \citet{Chabrier:2003} IMF and $Z$=0.008 metal 
abundance. Also, this allows truncated star formation history with 
constant star-formation in duration time of 10$^{6.5}$ to 10$^{11}$ 
yr, or delayed exponentially declining star-formation (SFR 
$\propto t \exp{(-t/\tau)}$) with log($\tau$/yr)=6.5--11. Then, 
star-formation history that gives better least chi square value is 
assigned. Here, log(age/yr) between 6 and 9 are given. Band 
photometries and errors of 
U,B,V,R,i,z,J,H,K,F606,814,125,140,160W from the 3D-HST catalogue 
are used, where \lya\ and \ha\ flux contaminations to B band and K 
band for LAEs and HAEs are corrected, respectively. $1\sigma$ 
errors of obtained values such as stellar mass are measured by 500 
Monte Carlo simulations attached with the {\sc fast} code. As a 
result, the median reduced chi-squares of the fitting results show 
1.18 and 1.83 in HAE and LAE samples, respectively. 

\begin{figure}
	\centering
	\includegraphics[width=75mm]{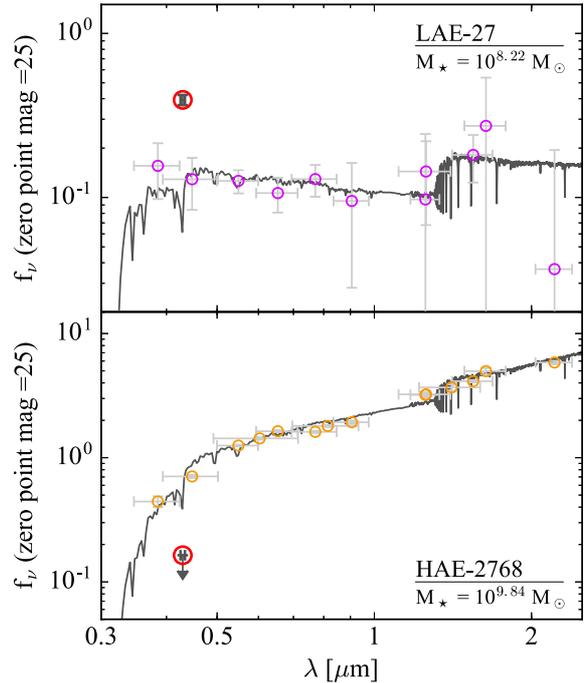}
    \caption{Examples of SED-fitting for LAEs (upper panel) and 
    HAEs (lower panel). Purple and orange circles indicate 
    observed broad-band flux densities, and large red circles 
    are flux densities of NB428. Black lines represent the 
    inferred spectra by SED-fitting.}
    \label{fig6}
\end{figure}

Fig.~\ref{fig6} represents examples of the SED-fitting for LAEs and
HAEs at $z=2.5$. We can see strong excess and deficit of NB flux
density in LAE and HAE, respectively. As compared to the HAE 
sample, LAEs tend to be faint at all broad-bands, and especially 
in the NIR regime F160W photometry is quite essential to determine 
their stellar masses. It should be noted again that emission line 
contaminations to broad-band photometries are negligible for our 
sample because F160W band does not include any strong emission 
line, and the fitting results remain consistent even if we apply 
restricted broad-band datasets which do not contain strong emission 
lines (\oii,\hb\,\oiii,\ha) to the SED-fitting. We also confirmed 
that measured stellar masses of all line emitters remain consistent 
regardless of assumption of star-formation history. 

In addition to the measurements of physical properties by the SED-fitting, 
we also estimate dust attenuation from the UV slope $\beta$ 
($f_\lambda\propto\lambda^\beta$) in accordance with the 
\citet{Meurer:1999, Calzetti:2000} relation. The UV slopes of all the emitters 
are calculated by error-weighted least chi square fitting based on 
V,R,i,F606W band photometries corresponding to $\lambda$=1500--2500 
\AA\ in the rest frame. We then estimate their 1 sigma errors by 1000 
Monte Carlo iterations. Recent studies tend to prefer to estimate the
stellar dust extinction from the UV slope since the dust 
reddening inferred from the whole SED-fitting would inevitably depend
on the stellar mass estimates \citep{Shivaei:2015}.
While the measured values of the UV slope for most of the
LAEs comfortably fall into the reasonable range inferred from the
\citet{Meurer:1999} relation within the margins of errors, there are
several LAEs which has very blue UV slopes ($\beta<-2.3$) deviating
from the relation (Table~\ref{tab2}). Such an extremely blue UV 
slope would require a large contribution of stellar 
populations of instantaneous starburst of the age $\lesssim10^7$ yr 
\citep{Leitherer:1995, Calzetti:2001} to the UV spectrum.
In this paper, we simply apply "zero" extinction for them.
On the other hand, SED-inferred extinction 
could be helpful since it may resolve varieties of stellar populations 
in such young LAEs. Therefore, we also check the results if we apply the 
SED-inferred extinctions, and those can be found in Appendix A. 

\begin{figure}
	\centering
	\includegraphics[width=75mm]{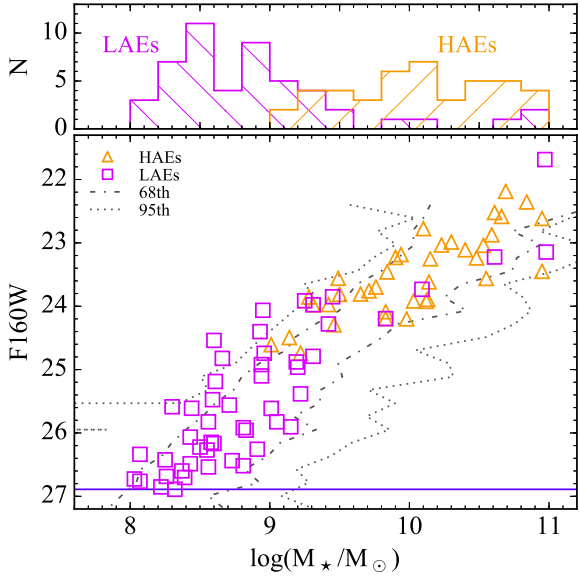}
    \caption{The upper histogram shows log stellar mass distributions 
    of LAEs (purple backslash) and HAEs (orange forward slash), and 
    the lower panel presents log stellar mass versus F160W magnitude. 
    The squares and the triangles are LAEs and HAE, respectively. The 
    horizontal line indicates 2$\sigma$ magnitude cut of F160W=26.89. 
    Dash-dot and dotted lines indicate the region enclosing 68\% and 
    95\% of photo-$z$ selected galaxies at $z$=2.3--2.8 for a given 
    F160W, based on the 3D-HST catalogue \citep{Skelton:2014}. 
    Our LAE sample is relatively complete above the stellar mass of 
    $\sim$10$^{8.5}$ \msun, while even lower mass LAEs should be 
    biased toward more active (bluer) star-forming galaxies.}
    \label{fig7}
\end{figure}

Figure~\ref{fig7} shows measured stellar mass distributions of LAEs 
and HAEs. LAEs tend to have lower stellar masses relative to \ha-flux 
limited sample. Although our sample size is not sufficient, our LAE 
sample in the target area would be relatively complete above the 
stellar mass of log(M$_\star$/M$_\odot$)$\sim$8.5 if the number of 
LAEs increase in smaller stellar masses. In the meantime, the lower 
panel in Fig.~\ref{fig7} shows F160W magnitudes as a function of log 
stellar mass of our all emitter sample. The figure also shows those 
enclosing 68\% and 95\% of galaxies at the photometric redshift of 
2.3--2.8 for a given F160W, which is taken from the 
\citet{Skelton:2014} catalogue. This suggests that the magnitude 
limit of F160W$<$26.89 can cover down to the stellar mass of 
$\sim$10$^{8.5}$ \msun. On the other hand, it also denotes that our 
selection should be biased toward galaxies with high mass-to-light 
ratios (i.e. more actively star-forming and bluer 
galaxies) in even lower mass regime $\lesssim$10$^{8.5}$ \msun. 
Similarly, we would have only active star-forming systems for the 
HAE sample below the stellar mass $\sim$10$^{10}$ \msun. 

The position of NB flux peaks and cross-matched ID number in the 
3D-HST catalogue \citep{Skelton:2014} and photometric redshifts by 
\citet{Momcheva:2016} for all LAEs can be found in Table~\ref{tab2}. 
This also describes the values and 1 sigma errors of those \lya\ 
luminosities, \lya\ equivalent widths in the rest frame, UV slopes, 
stellar masses, dust-corrected SFRs based on E(B$-$V) from UV slope 
and SED-fitting, and effective radii from \citet{Wel:2012}.

\begin{table*}
	\centering
	\caption{Final LAE catalogue at $z=2.5$. The first column shows 
    identification number of LAEs. The second and fifth columns indicate 
    identification number in the 3D-HST UDS catalogue \citep{Skelton:2014} 
    and photometric redshifts ($z_\mathrm{best}$) and those 68th 
    percentiles from \citet{Momcheva:2016}, based on the 3D-HST Treasury 
    Survey. The second and third show peak position of NB flux. The sixth 
    to ninth represent values and 1 sigma errors of \lya\ luminosity (E41 
    erg/s), EW of \lya\ line emission in the rest frame (\AA), UV slope 
    $\beta$, log stellar mass. 
    The 10th and the 11th are SFRs from UV luminosity (\msun/yr) where 
    those dust extinctions are corrected, following UV slope and 
    SED-fitting, respectively. The 12th are effective radii derived from 
    {\sc galfit} distributed by \citealt{Wel:2012} (flags indicate the 
    fitting results defined by their paper; $s$=suspicious, $b$=bad). 
    Full table is available as online material.}
	\label{tab2}
	\begin{tabular}{rrrrrrrrrrrr} 
		\hline
		$^1$ID & $^2$ID$_\mathrm{HST}$ & $^3$R.A.$_\mathrm{NB}$ & $^4$Dec.$_\mathrm{NB}$ 
        & $^5$$z_\mathrm{best}$ & $^6$L$_\mathrm{E41}$\hspace{0.2cm} & $^7$EW & $^8$$\beta$\hspace{0.4cm} & 
        $^9$logM$_\star$ \hspace{0.1cm} & $^{10}$SFR$_\mathrm{UV_\beta}$ & $^{11}$SFR$_\mathrm{UV_{sed}}$ &
        $^{12}$R$_\mathrm{e}$\hspace{0.3cm}  \\
		\hline
   1 &  5917 & 34.54872 & $-$5.25659 & 1.71 &  3.4$\pm$1.4 &  23$\pm$5  & $-$2.71$\pm$0.39 &  8.73$_{-1.02}^{+0.18}$ &  1.0$_{-0.1}^{+4.9}$   &  1.0$_{-0.1}^{+1.2}$   & $^b$0.38$\pm$0.42 \\
   7 & 10425 & 34.56473 & $-$5.24084 & 2.58 & 12.7$\pm$2.4 &  21$\pm$4  & $-$1.74$\pm$0.03 &  9.20$_{-0.80}^{+0.17}$ &  5.3$_{-3.3}^{+17.4}$  &  6.8$_{-4.8}^{+6.3}$   &     1.12$\pm$0.36 \\
   9 & 14560 & 34.55527 & $-$5.22678 & 2.04 & 13.6$\pm$2.2 &  39$\pm$5  & $-$2.77$\pm$0.16 &  8.59$_{-0.58}^{+0.50}$ &  2.4$_{-0.2}^{+10.8}$  &  2.4$_{-0.2}^{+4.6}$   &     1.52$\pm$0.85 \\
  12 & 18325 & 34.56935 & $-$5.21441 & 1.42 & 20.8$\pm$1.5 & 140$\pm$24 & $-$2.85$\pm$0.35 &  8.60$_{-0.80}^{+0.26}$ &  0.8$_{-0.2}^{+4.0}$   &  0.8$_{-0.2}^{+1.5}$   &     0.75$\pm$0.54 \\
  15 & 18990 & 34.55072 & $-$5.21225 & 1.11 &  9.6$\pm$1.7 &  21$\pm$4  & $-$1.54$\pm$0.25 &  8.30$_{-0.22}^{+0.68}$ &  7.9$_{-5.9}^{+26.6}$  &  7.0$_{-5.0}^{+5.7}$   &     2.35$\pm$1.60 \\
  22 & 37097 & 34.58892 & $-$5.15123 & 2.23 & 46.0$\pm$1.9 &  15$\pm$0  & $-$2.40$\pm$0.21 &  9.45$_{-0.06}^{+0.22}$ &  8.3$_{-0.1}^{+37.6}$  &  8.3$_{-0.1}^{+5.9}$   &     1.69$\pm$0.12 \\
  27 & 34511 & 34.51098 & $-$5.16014 & 1.56 &  5.6$\pm$1.4 &  41$\pm$9  & $-$2.14$\pm$0.38 &  8.22$_{-1.09}^{+0.34}$ &  0.6$_{-0.2}^{+2.9}$   &  0.5$_{-0.1}^{+0.6}$   & $^b$0.62$\pm$0.84 \\
  28 & 28289 & 34.56316 & $-$5.18097 & 2.41 & 48.6$\pm$2.9 &  16$\pm$1  & $-$2.28$\pm$0.23 &  8.60$_{-0.05}^{+0.68}$ &  5.0$_{-0.1}^{+22.8}$  & 15.7$_{-10.9}^{+5.4}$  &     1.28$\pm$0.16 \\
  29 & 32164 & 34.50610 & $-$5.16808 & 2.54 & 18.4$\pm$1.8 &  20$\pm$2  & $-$1.91$\pm$0.20 &  8.66$_{-0.00}^{+0.08}$ &  6.8$_{-3.1}^{+26.6}$  & 18.8$_{-2.6}^{+4.4}$   &     0.61$\pm$0.19 \\
  30 & 30895 & 34.50703 & $-$5.17256 & 2.33 & 49.0$\pm$1.5 &  38$\pm$1  & $-$2.14$\pm$0.22 &  9.42$_{-0.16}^{+0.16}$ &  6.3$_{-1.0}^{+27.6}$  & 10.7$_{-5.4}^{+3.2}$   &     1.43$\pm$0.14 \\
  $^m$31 & 30133 & 34.54215 & $-$5.17551 & 1.84 &  136$\pm$1.4 &  48$\pm$0  & $-$1.58$\pm$0.41 & 10.97$_{-0.31}^{+0.19}$ & 52.0$_{-36.5}^{+193}$  &  628$_{-612}^{+825}$   &     1.11$\pm$0.05 \\
  35 &  3182 & 34.48001 & $-$5.26605 & 2.51 & 11.7$\pm$1.3 &  68$\pm$13 & $-$0.41$\pm$0.09 & 10.61$_{-0.08}^{+0.07}$ & 26.2$_{-24.9}^{+24.9}$ & 22.5$_{-21.7}^{+24.5}$ &     1.23$\pm$0.06 \\
  36 &  4309 & 34.44040 & $-$5.26320 & 2.51 &  7.7$\pm$1.5 &  77$\pm$24 & $-$0.36$\pm$0.26 & 10.98$_{-0.22}^{+0.10}$ & 30.9$_{-30.0}^{+36.1}$ & 24.1$_{-23.3}^{+43.5}$ &     2.97$\pm$0.13 \\
  41 & 10396 & 34.43763 & $-$5.24091 & 1.90 & 10.9$\pm$1.6 & 173$\pm$33 & $-$2.18$\pm$0.50 &  8.81$_{-0.38}^{+0.17}$ &  0.4$_{-0.1}^{+1.9}$   &  0.4$_{-0.1}^{+0.6}$   &     1.10$\pm$0.73 \\
  47 & 20895 & 34.47444 & $-$5.20549 & 2.43 &  5.5$\pm$1.6 &  30$\pm$7  & $-$1.74$\pm$0.03 &  8.50$_{-0.28}^{+0.42}$ &  1.6$_{-1.0}^{+5.2}$   &  0.6$_{-0.1}^{+1.7}$   & $^b$1.35$\pm$0.62 \\
  56 & 27668 & 34.47025 & $-$5.18285 & 0.56 &  6.2$\pm$1.4 &  34$\pm$8  & $-$2.78$\pm$0.31 &  8.07$_{-0.60}^{+0.40}$ &  1.5$_{-0.1}^{+7.3}$   &  1.9$_{-0.5}^{+3.7}$   &     2.00$\pm$1.26 \\
  57 & 27420 & 34.42780 & $-$5.18368 & 2.40 & 11.7$\pm$1.6 &  51$\pm$11 & $-$2.00$\pm$0.40 &  9.15$_{-0.93}^{+0.16}$ &  1.8$_{-0.8}^{+8.1}$   &  3.0$_{-1.9}^{+4.9}$   & $^b$1.88$\pm$1.12 \\
  58 & 26371 & 34.48166 & $-$5.18751 & 2.54 & 23.6$\pm$2.4 &  19$\pm$3  & $-$1.75$\pm$0.27 &  8.61$_{-0.13}^{+0.54}$ &  5.2$_{-3.1}^{+19.5}$  & 16.9$_{-9.8}^{+13.6}$  & $^b$1.82$\pm$0.29 \\
  60 & 24704 & 34.46305 & $-$5.19313 & 2.66 & 22.6$\pm$1.4 &  15$\pm$1  & $-$1.77$\pm$0.31 &  8.95$_{-0.00}^{+0.65}$ & 11.1$_{-6.5}^{+42.9}$  & 47.2$_{-42.6}^{+13.1}$ &     0.87$\pm$0.11 \\
  61 & 24397 & 34.47254 & $-$5.19386 & 2.64 &  9.8$\pm$1.9 & 223$\pm$90 & $-$2.39$\pm$0.33 &  8.39$_{-0.26}^{+0.38}$ &  0.2$_{-0.0}^{+1.0}$   &  0.2$_{-0.0}^{+1.2}$   & $^b$0.98$\pm$0.63 \\
  63 &  5744 & 34.40507 & $-$5.25718 & 2.23 & 24.4$\pm$2.5 &  22$\pm$2  & $-$2.09$\pm$0.25 &  8.94$_{-0.30}^{+0.16}$ &  3.6$_{-0.9}^{+15.6}$  &  5.5$_{-2.8}^{+3.2}$   &     0.78$\pm$0.23 \\
  64 & 11017 & 34.32240 & $-$5.23897 & 2.20 & 21.2$\pm$1.7 &  50$\pm$4  & $-$2.26$\pm$0.36 &  8.43$_{-0.58}^{+0.36}$ &  1.3$_{-0.1}^{+6.4}$   &  2.7$_{-1.4}^{+2.0}$   & $^b$1.06$\pm$0.79 \\
  67 & 12523 & 34.33171 & $-$5.23398 & 2.51 & 25.1$\pm$1.6 &  31$\pm$3  & $-$2.26$\pm$0.22 &  8.44$_{-0.18}^{+0.43}$ &  2.0$_{-0.1}^{+9.1}$   & 15.8$_{-13.9}^{+19.0}$ & $^b$0.87$\pm$0.28 \\
  73 & 18238 & 34.32620 & $-$5.21464 & 1.89 &  2.8$\pm$1.3 &  41$\pm$12 & $-$1.75$\pm$0.56 &  8.55$_{-0.69}^{+0.41}$ &  3.9$_{-2.5}^{+16.6}$  &  3.1$_{-1.8}^{+6.5}$   & $^b$142$\pm$153   \\
  74 & 18921 & 34.34030 & $-$5.21269 & 2.00 &  5.7$\pm$1.4 &  24$\pm$3  & $-$2.36$\pm$0.31 &  9.22$_{-0.35}^{+0.06}$ &  1.9$_{-0.1}^{+8.8}$   &  1.9$_{-0.1}^{+1.4}$   & $^b$2.59$\pm$0.66 \\
  76 & 39205 & 34.34621 & $-$5.14343 & 2.00 & 41.4$\pm$1.4 &  97$\pm$7  & $-$2.21$\pm$0.19 &  8.56$_{-0.58}^{+0.20}$ &  1.5$_{-0.1}^{+6.7}$   &  4.7$_{-3.3}^{+5.1}$   &     0.67$\pm$0.19 \\
  78 & 37204 & 34.37653 & $-$5.15058 & 2.37 & 15.9$\pm$1.7 &  51$\pm$7  & $-$2.11$\pm$0.18 &  9.01$_{-0.73}^{+0.24}$ &  1.3$_{-0.3}^{+5.4}$   &  5.2$_{-4.2}^{+5.8}$   &     1.16$\pm$0.39 \\
  79 & 22942 & 34.39716 & $-$5.19887 & 2.56 & 17.8$\pm$1.3 &  38$\pm$3  & $-$1.88$\pm$0.28 &  8.94$_{-0.33}^{+0.32}$ &  4.9$_{-2.4}^{+19.6}$  &  3.2$_{-0.8}^{+6.4}$   &     1.08$\pm$0.24 \\
  84 & 29768 & 34.31855 & $-$5.17587 & 1.41 & 14.3$\pm$1.6 & 112$\pm$20 & $-$2.86$\pm$0.40 &  8.37$_{-0.43}^{+0.19}$ &  0.6$_{-0.1}^{+3.0}$   &  0.6$_{-0.1}^{+0.4}$   & $^b$1.11$\pm$1.03 \\
  $^h$87 & 27615 & 34.36791 & $-$5.18348 & 2.39 & 83.7$\pm$2.3 &  27$\pm$0  & $-$2.26$\pm$0.25 &  9.31$_{-0.15}^{+0.14}$ &  7.2$_{-0.1}^{+33.3}$  &  9.1$_{-2.0}^{+6.3}$   & $^s$2.59$\pm$0.14 \\
  90 & 24644 & 34.34509 & $-$5.19317 & 1.71 & 20.0$\pm$1.3 & 139$\pm$26 & $-$3.17$\pm$0.23 &  8.32$_{-0.54}^{+0.21}$ &  0.5$_{-0.1}^{+2.3}$   &  0.5$_{-0.1}^{+0.5}$   & $^b$1.51$\pm$1.46 \\
  91 & 24422 & 34.38580 & $-$5.19380 & 1.68 & 27.7$\pm$2.0 &  56$\pm$5  & $-$2.47$\pm$0.20 &  8.58$_{-0.91}^{+0.26}$ &  1.6$_{-0.1}^{+7.3}$   &  1.6$_{-0.1}^{+1.5}$   & $^b$2.24$\pm$1.31 \\
  92 & 22925 & 34.34627 & $-$5.19876 & 1.53 & 12.8$\pm$1.8 &  41$\pm$9  & $-$1.95$\pm$0.53 &  8.25$_{-0.61}^{+0.36}$ &  2.1$_{-1.0}^{+9.6}$   &  1.2$_{-0.2}^{+1.1}$   & $^b$0.48$\pm$0.42 \\
  93 &  2452 & 34.22792 & $-$5.26813 & 1.72 &  5.4$\pm$0.9 &  48$\pm$15 & $-$2.74$\pm$0.42 &  8.56$_{-0.77}^{+0.38}$ &  0.7$_{-0.1}^{+3.5}$   &  0.7$_{-0.1}^{+1.7}$   &     0.51$\pm$0.42 \\
  94 &  2967 & 34.23742 & $-$5.26633 & 1.91 &  7.8$\pm$1.0 &  36$\pm$10 & $-$1.87$\pm$0.22 &  8.91$_{-0.93}^{+0.15}$ &  1.1$_{-0.7}^{+4.3}$   &  1.4$_{-1.0}^{+2.0}$   &     0.60$\pm$0.40 \\
  95 &  3013 & 34.22358 & $-$5.26614 & 1.00 &  6.8$\pm$1.7 &  21$\pm$4  & $-$0.60$\pm$0.38 &  8.71$_{-0.29}^{+0.34}$ & 24.1$_{-23.0}^{+43.9}$ & 46.1$_{-45.1}^{+28.1}$ &     1.21$\pm$0.30 \\
 100 &  8055 & 34.28063 & $-$5.24924 & 2.39 & 15.7$\pm$2.2 &  24$\pm$2  & $-$2.59$\pm$0.12 &  8.96$_{-0.09}^{+0.27}$ &  2.9$_{-0.1}^{+12.6}$  &  2.9$_{-0.1}^{+1.7}$   &     4.85$\pm$1.48 \\
 101 &  8542 & 34.29174 & $-$5.24732 & 1.74 & 17.5$\pm$2.1 &  43$\pm$8  & $-$1.82$\pm$0.44 &  8.43$_{-0.61}^{+0.17}$ &  2.1$_{-1.2}^{+9.0}$   & 20.1$_{-19.2}^{+9.9}$  &     0.36$\pm$0.19 \\
 102 &  9191 & 34.22307 & $-$5.24571 & 1.40 &  4.3$\pm$1.6 &  17$\pm$3  & $-$0.85$\pm$0.24 & 10.09$_{-0.00}^{+0.27}$ & 22.9$_{-21.3}^{+46.9}$ &  2.2$_{-0.5}^{+8.0}$   &     6.24$\pm$0.54 \\
 108 & 12330 & 34.30823 & $-$5.23463 & 1.67 & 28.2$\pm$2.3 &  71$\pm$6  & $-$2.28$\pm$0.41 &  8.83$_{-0.17}^{+0.09}$ &  1.3$_{-0.1}^{+6.4}$   &  1.3$_{-0.1}^{+0.3}$   &     0.51$\pm$0.24 \\
 110 & 18016 & 34.29846 & $-$5.21541 & 1.39 & 12.1$\pm$1.9 &  24$\pm$4  & $-$2.88$\pm$0.05 &  9.19$_{-0.35}^{+0.15}$ &  1.2$_{-0.1}^{+5.2}$   &  4.9$_{-3.8}^{+17.5}$  &     0.71$\pm$0.13 \\
 111 & 18221 & 34.27096 & $-$5.21479 & 1.39 &  7.3$\pm$1.1 &  87$\pm$23 & $-$2.15$\pm$0.38 &  8.26$_{-0.60}^{+0.34}$ &  0.5$_{-0.2}^{+2.5}$   &  0.5$_{-0.1}^{+1.4}$   & $^b$0.80$\pm$0.45 \\
 112 & 18947 & 34.22298 & $-$5.21232 & 1.62 &  7.3$\pm$1.4 &  23$\pm$4  & $-$2.54$\pm$0.33 &  8.81$_{-0.37}^{+0.29}$ &  1.1$_{-0.1}^{+5.3}$   &  2.2$_{-1.2}^{+3.4}$   &     1.13$\pm$0.37 \\
 $^d$114 & 29191 & 34.25741 & $-$5.17772 & 1.04 & 11.0$\pm$1.4 & 148$\pm$53 & $-$2.46$\pm$0.33 &  9.05$_{-0.48}^{+0.23}$ &  0.2$_{-0.2}^{+1.1}$   &  0.2$_{-0.2}^{+1.3}$   & $^b$0.98$\pm$0.69 \\
 117 & 34684 & 34.24193 & $-$5.15978 & 2.50 & 34.2$\pm$1.9 &  23$\pm$1  & $-$1.70$\pm$0.18 &  9.25$_{-0.00}^{+0.07}$ & 11.7$_{-7.4}^{+40.8}$  & 69.1$_{-2.0}^{+16.0}$  &     1.10$\pm$0.09 \\
 $^h$118 & 33079 & 34.24434 & $-$5.16520 & 2.40 & 29.0$\pm$1.4 &  36$\pm$2  & $-$1.82$\pm$0.14 &  9.83$_{-0.02}^{+0.07}$ &  5.9$_{-3.2}^{+21.6}$  &  4.4$_{-0.5}^{+3.6}$   &     0.42$\pm$0.06 \\
 119 & 32712 & 34.23558 & $-$5.16694 & 1.04 & 16.3$\pm$2.9 &  36$\pm$5  & $-$0.67$\pm$0.47 &  8.93$_{-0.42}^{+0.47}$ & 41.6$_{-39.5}^{+88.0}$ & 14.4$_{-12.4}^{+28.3}$ &     5.82$\pm$1.34 \\
 121 & 28511 & 34.24313 & $-$5.18004 & 1.70 & 13.0$\pm$1.8 &  54$\pm$10 & $-$1.83$\pm$0.41 &  8.03$_{-0.58}^{+0.48}$ &  1.9$_{-1.1}^{+7.9}$   &  0.9$_{-0.1}^{+2.2}$   & $^b$0.14$\pm$0.19 \\
 122 & 27412 & 34.26027 & $-$5.18367 & 1.51 & 13.0$\pm$1.4 &  95$\pm$23 & $-$1.50$\pm$0.24 &  8.07$_{-0.50}^{+0.59}$ &  1.7$_{-1.4}^{+5.6}$   &  1.8$_{-1.4}^{+4.6}$   & $^b$1.85$\pm$1.95 \\
 123 & 24058 & 34.27530 & $-$5.19516 & 2.23 & 14.5$\pm$1.9 &  21$\pm$3  & $-$0.42$\pm$0.29 &  9.31$_{-0.32}^{+0.11}$ & 54.7$_{-52.9}^{+72.8}$ &  2.4$_{-0.6}^{+2.4}$   & $^b$2.83$\pm$0.36 \\
		\hline \\
        \multicolumn{12}{l}{$^h$ \ha\ emitters at $z$=2.5 selected by \citet{Tadaki:2013, Tadaki:2014}.} \\
        \multicolumn{12}{l}{$^m$ MIPS 24$\micron$m sources found by SpUDS \citep{Dunlop:2007} overlap those objects within 3 arcsec radius.} \\
        \multicolumn{12}{l}{$^x$ All LAEs have no X-ray source detected by Subaru/XMM-Newton Deep Survey \citep{Ueda:2008} within 3 arcsec radius.} \\
        \multicolumn{12}{l}{$^d$ affected by diffraction spikes in HST images.}
	\end{tabular}
\end{table*}

\section{Results} 

\subsection{Main sequence}

We now study star-forming activities of the LAEs as well as those of 
HAEs at $z=2.53$. We employ UV luminosities measured from V-band 
magnitudes, which correspond to the UV flux densities at $\sim$1600 
\AA\ in the rest frame. The deep \lya\ data and multi-band datasets 
allow us to reach and investigate the low-mass end of the main 
sequence at $z$=2.5. As mentioned in the previous section, LAEs below 
the stellar mass of log(M$_\star$/M$_\odot$)$=$8.5 will be weighted 
toward more active star-forming systems.

Since \lya\ emission line is located at UV wavelength range unlike 
rest-frame optical emission lines like \ha, the selection effect is 
critical when one studies star-forming activities in LAEs. This is 
especially true for narrow-band selected LAEs, which ought to be 
biased to LAEs with high EWs. In the case of \lya\ emission line, 
EW$_\mathrm{Ly\alpha}$ is always lower than $\sim$240\AA\ according to 
the stellar population model 
unless we take into account the contributions from 
Pop-III type objects, accreting binary stars, and/or rapidly rotating 
massive stars 
\citep{Schaerer:2003,Malhotra:2002,Kashikawa:2012,Trainor:2016}.
Thus the shallow spectroscopic survey of LAEs is likely to be biased 
toward LAEs with luminous UV flux densities 
($\gtrsim L_\mathrm{Ly\alpha}$/240 erg s$^{-1}$\AA$^{-1}$) as well. 
Assuming the typical UV slope of $\beta=-2.2$ seen in our low-mass LAEs 
($<$10$^{10}$ \msun), the depths of our narrow-band and B-band data 
(NB$_{5\sigma}$=26.55 and B$_{5\sigma}$=27.89, respectively) 
can capture LAEs with EW$_\mathrm{Ly\alpha}\geq15$, 
20, and 50 \AA\ down to SFRs of 0.8, 0.5, and 0.3 \msun/yr, 
respectively. Therefore, our data are able to reach typical 
star-forming activities of galaxies with the stellar masses down to 
$\sim10^8$ \msun\ according to the extrapolated main sequence of 
massive star-forming galaxies at similar redshfits 
\citep{Skelton:2014,Shivaei:2015,Tomczak:2016}. This is a great 
advancement compared to other similar studies such as 
\citet{Vargas:2014,Hagen:2014,Hagen:2016,Oteo:2015} which are limited 
towards LAEs with higher SFRs due to larger line luminosity limits.

\begin{figure*}
	\includegraphics[width=165mm]{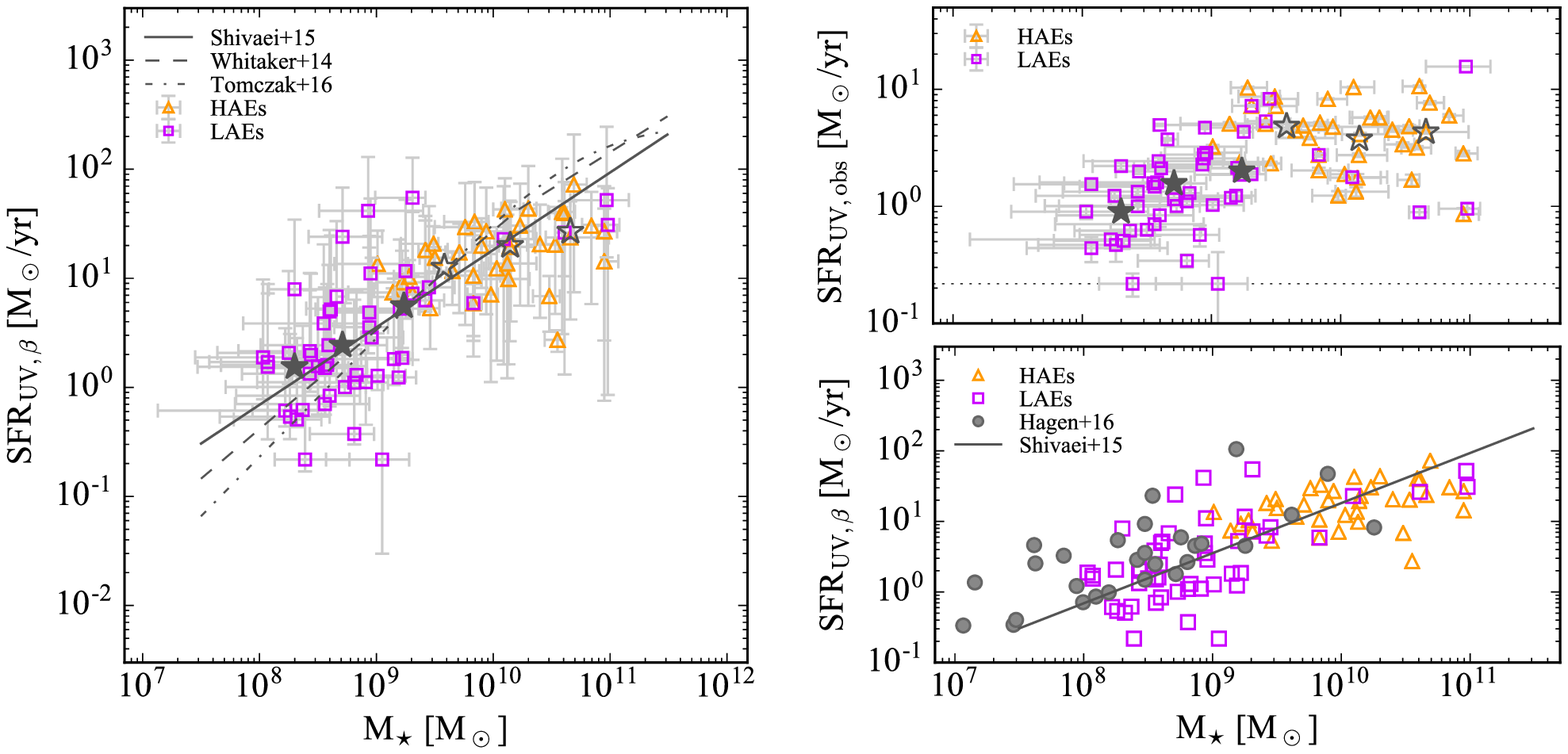}
    \caption{(a) The left panel shows the stellar mass versus 
    dust-corrected SFR, SFR$\mathrm{UV,\beta}$. Extinction correction 
    is based on UV slope. (b) The upper right panel presents 
    dust-uncorrected SFR, SFR$_\mathrm{UV,obs}$, and (c) the bottom 
    right panel shows the same as (a), but the result by 
    \citet{Hagen:2016} are also plotted for comparison (grey filled 
    circles). Orange triangles and purple squares represent HAEs and 
    LAEs at $z=2.5$ (this work). The observed UV luminosities are 
    measured from V-band, and the $2\sigma$ limiting magnitude in V-band 
    (V=28.19) is shown by the dotted horizontal line in the panel (c). 
    The black filled stars in panels (a) and (b) indicate the median SFRs 
    of the LAEs in the stellar mass bins in logarithmic scale of $<$8.5, 
    8.5--9.0, and 9.0--10.0, while the open stars present the median SFRs 
    of HAEs at the log stellar masses of $<$10.0, 10.0--10.5, and $>$10.5. 
    In the panel (a), dashed, dot-dash, and solid lines respectively show 
    the main sequences of the star-forming galaxies at $z$=2.0--2.5 
    reported by \citet{Whitaker:2014b}, the one at $z$=2.5 obtained by 
    the prescription in \citet{Tomczak:2016}, and the one at $z$=2.0--2.6 
    by \citet{Shivaei:2015}. The SFRs are all obtained based on UV+IR 
    luminosities in  \citet{Whitaker:2014b,Tomczak:2016}, and the same 
    manner as our measurement in \citet{Shivaei:2015}, respectively. 
    Since the main sequences for comparison are derived only for massive 
    galaxies above $1\times10^{9}$ or $1\times10^{10}$ \msun\ at 
    $z\sim2.5$, we just linearly extrapolate the relations to the lower 
    mass regime. The measurements of SFRs (and dust correction) using 
    other methods are also presented in the Appendix.}
    \label{fig8}
\end{figure*}

Figure~\ref{fig8}b presents stellar masses versus dust-uncorrected 
SFRs of the LAEs and HAEs. We measure the SFRs based on the 
\citet{Kennicutt:1998} conversion assuming the \citet{Chabrier:2003} 
IMF. The $2\sigma$ limiting magnitude at V-band is 28.19 mag which 
corresponds to the SFR limit of 0.2 \msun/yr. The $2\sigma$ flux 
densities are assumed for two emitters that are fainter than this 
threshold. The median values are calculated in each mass bin 
corresponding to log stellar mass $<$8.5, 8.5--9.0, and 9.0--10.0 for 
LAEs, and log stellar mass $<$10.0, 10.0--10.5, and $>$10.5 for HAEs, 
respectively. Since more massive galaxies (HAEs) tend to be dustier 
and/or beginning to be quenched, the slope of the mass vs.\ UV 
luminosity relation turns over at the massive end roughly above 
$1\times10^{10}$ \msun. On the other hand, most of the LAEs contain 
only a small amount of dust and tend to be less massive. Therefore, 
they show a positive correlation between stellar masses and observed 
UV luminosities even without a dust correction. 

We then estimate the dust-corrected SFRs of our sample and make an 
improved mass--SFR relation (star-forming main sequence) 
\citep{Daddi:2007a,Elbaz:2007,Noeske:2007}, which is compared to 
earlier but recent measurements in the literature 
\citep{Speagle:2014,Whitaker:2014b,Shivaei:2015,Tomczak:2016}.
The extinction levels are obtained based on the UV slope 
(Fig.~\ref{fig8}a) or the SED-fitting (Appendix A) as mentioned in 
\S2.5. The SED-inferred SFRs are also shown in the Appendix. Before a 
comparison is made, several remarks should be noted. The main sequences 
reported by \citep{Whitaker:2014b,Tomczak:2016} are obtained by UV+IR 
luminosities from the stacked data of photo-$z$ selected galaxies. On 
the other hand, \citet{Shivaei:2015} reported the main sequence of 
spectroscopically confirmed galaxies at $z=$2--2.6 selected by 
F160W-band magnitude limit ($<25$ mag), and they have studied the main 
sequence with different varieties of SFR measurements. In 
Fig.~\ref{fig8}a,c, the main sequence from \citet{Shivaei:2015} is 
based on SFR from UV with dust correction by UV slope, which is the 
same way as our estimation here. The choice of SFR calibration and 
this impact on measurements of the main sequence is beyond the scope 
of this paper. However, it is important to check and discuss this 
bias. Thus this paper also shows the same diagrams from different 
SFR measurements in the Appendix A. 

Although our LAEs are widely spread in stellar mass, the median 
values follow neatly along the linearly extrapolated lines of the 
main sequences of \citet{Whitaker:2014b,Shivaei:2015,Tomczak:2016} 
as shown in Fig.~\ref{fig8}a. The measured SFRs and stellar masses 
of individual LAEs are given in Table~\ref{tab2}, and the median 
values plotted in the Fig~\ref{fig8} are also listed in 
Table~\ref{tab3}. Two HAEs classified also as LAEs just overlap each 
other in the figures. 
It should be noted that this paper has not used seven heavily obscured 
HAEs missing B-band detections to be consistent with the LAE selection 
(see \S2.5). This selection effect underestimates the median values of 
HAEs at the massive end ($>10^{10.5}$), which include five out of 
those in the Fig.~\ref{fig8}. 

In Fig.~\ref{fig8}c, the distribution of our LAE sample on the 
mass--SFR diagram is clearly different from that of LAEs at 
$z$=1.9--2.3 explored by \citet{Hagen:2016}. Such inconsistency is 
likely to be caused by the shallower data in \citet{Hagen:2016}. 
However, we should notify that \citet{Hagen:2016} have confirmed 
the same mass--SFR distribution between LAEs and their `control' 
non-LAE sample in their paper. In that sense, our result is consistent 
with the report by \citet{Hagen:2016}, and indeed the mass--SFR 
distribution in their LAEs do not deviate from those of relatively 
active star-forming LAEs in our sample. 

Our data now confirm that the low-mass LAEs are located along the 
expected, extrapolated line of the main sequence of normal massive 
star-forming galaxies towards lower stellar masses.
However, it should be noted that we do not know whether the LAEs are 
the representative populations of the low-mass systems at this redshift 
because only $\sim10\%$ of star-forming galaxies would be selected as 
LAEs \citep{Hayes:2010,Cassata:2015,Matthee:2016,Hathi:2016,Sobral:2016}. 
Moreover, we do not have a comparison sample (i.e.\ non-\lya-emitting 
galaxies) at the mass regime below $1\times10^9$ \msun.

On the other hand, such a similarity of star-forming activities 
between LAEs and normal star-forming galaxies may not apply to more 
massive LAEs. Four LAEs with the stellar mass of $>10^{9.5}$ solar 
mass tend to be located below the main sequence, which is notable 
in SFR$_\mathrm{UV,sed}$ and SFR$_\mathrm{SED}$ (see Appendix). 
Such less-active LAEs are also discovered by \citet{Taniguchi:2015}. 
We discuss these LAEs in \S4.2.

\subsection{Mass-size relation}

\begin{figure}
	\includegraphics[width=80mm]{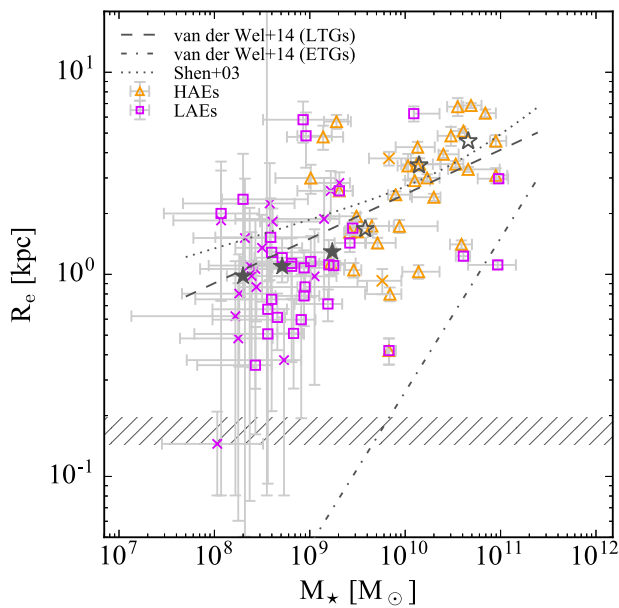}
    \caption{Stellar masses versus effective radii R$_\mathrm{e}$ 
    of LAEs and HAEs at $z=2.5$. The symbols are the same as in 
    Fig.~\ref{fig8}, but the objects whose effective radii are 
    not fitted well and marked as `bad fit' in the catalogue of 
    \citet{Wel:2012}, are shown by the cross marks. The hatched region 
    corresponds to the size of bright stars ($\pm1\sigma$).
    The Dashed and dotted lines indicate the mass-size 
    relations of late-type and early-type galaxies at $z$=2.0--2.5 
    reported by \citet{Wel:2014}, respectively. Dotted curve shows 
    the local mass--size relation of late-type galaxies 
    \citep{Shen:2003}.}
    \label{fig9}
\end{figure}

Typical LAEs follow the extrapolated line of the main sequence of 
massive star-forming galaxies as presented in the previous section. 
How about their stellar angular sizes? In this section, we explore 
the positions of the LAEs on the mass--size diagram. We use the 
effective radii on the semi-major axis which are measured from the
HST F160W image and listed in the catalogue of \citet{Wel:2012, Wel:2014}.
They use the software, {\sc galfit} \citep{Peng.CY:2010} combined with 
{\sc galapagos} \citep{Barden:2012} in order to measure the precise
background levels. Details of the fitting procedure and associated error 
estimation can be found in their paper. Since their catalogue uses the 
same identification numbers as those in the 3D-HST catalogue 
\citep{Skelton:2014}, we were easily able to obtain the stellar sizes 
for all the LAEs, by cross-matching the catalogues. 

The result is shown in Figure~\ref{fig9}. Because of the faintness 
of the sources, no reliable fitting results can be obtained for 20
out of all the 50 LAEs in the F160W image, and they are marked as 
`bad fit' or `suspicious fit' in the \citet{Wel:2012} catalogue. It is 
quite hard to measure the sizes of the low-mass LAEs below $1\times10^9$ 
\msun, in particular for diffuse (lower surface brightness) LAEs. 
Even if the fits are reasonably good, LAEs tend to be faint in F160W 
images, and their morphologies are largely unresolved. In this section, 
we only employ the effective radii in their catalogue. However, the size 
measurements of faint objects at F160W/IR image are uncertain (see also 
\citealt{Shibuya:2015}), and the results should be seen with this 
caution in mind. 

Although the size measurements of the LAEs tend to have large 
error-bars compared to the HAEs at the same redshift, the LAEs seem to 
lie on the area where the mass--size relation of star-forming galaxies 
at $z$=2--2.5 by \citet{Wel:2014} is extrapolated to the lower mass 
regime. We do not find any statistical difference in their angular 
sizes with respect to the HAEs along the mass-size relation. We also 
check and confirm that the distribution of our LAE sample on the 
mass--size diagram agrees well with that of LAEs at $z\sim2$ by 
\citet{Hagen:2016}, which also reports a consistent mass--size 
distribution between non-LAEs and LAEs at the same stellar mass range. 
We also confirm that the size of our LAEs are consistent with those of 
LAEs at $z=2.1$ and 3.1 by \citet{Bond:2012}, based on the 
measurements with the SExtractor \citep{Skelton:2014}.

Our result suggests that the stellar sizes of low-mass LAEs are not 
biased, and they are seen as just normal low-mass star-forming 
galaxies. However, we find four compact galaxies for their masses 
among five massive LAEs with log(M$_\star$/M$_\odot$)$>9.5$. They are 
actually located near the mass--size relation of early-type galaxies 
at $z$=2--2.5 by \citet{Wel:2014}. Although four objects without 
spectroscopic redshifts are certainly insufficient to conclude the 
compact nature of massive \lya\ emitters, they are good candidates for 
compact star-forming galaxies at $z>2$ as progenitors of massive 
compact quiescent galaxies at later epochs 
\citep{Barro:2013,Tadaki:2014}. We confer these objects in the 
discussion section \S4.2.

The median values and the 1 $\sigma$ scatter values of R$_\mathrm{e}$ 
in different stellar mass bins for our LAE sample seen in the 
Fig.~\ref{fig9} are listed in the Table~\ref{tab3}.

\subsection{Narrow-band absorbers in HAEs at z=2.5}

\begin{figure*}
	\includegraphics[width=130mm]{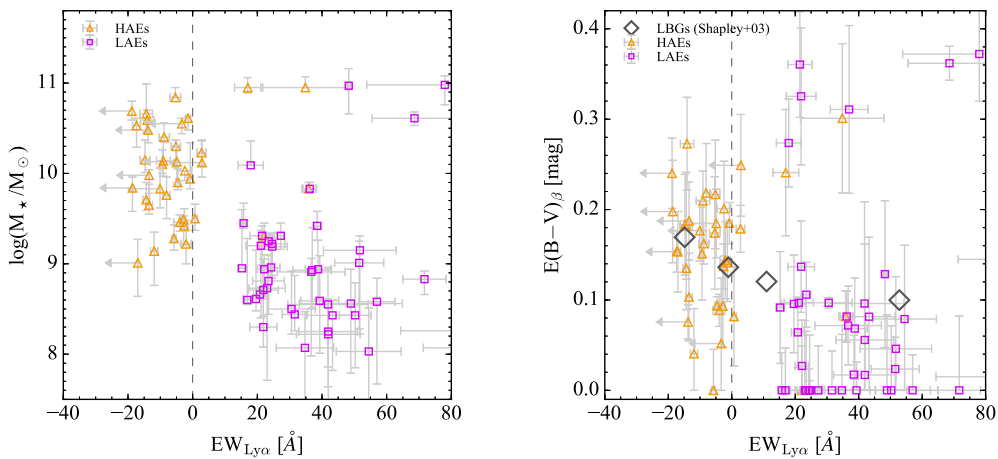}
    \caption{EW$_\mathrm{Ly\alpha}$ vs. log stellar mass (a; left), 
    E(B$-$V) derived from UV slope (b; right). The symbols are the 
    same as shown in Fig.~\ref{fig8}. The black diamonds in the right 
    panel represent the results of composite spectra of Lyman break 
    galaxies at $z\sim$3 reported by \citet{Shapley:2003}. Dashed 
    vertical lines indicate EW$_\mathrm{Ly\alpha}=0$ \AA. To 
    facilitate visualization, the figures only show the range of 
    EW$_\mathrm{Ly\alpha}$ between $-$40 and 80 \AA.}
    \label{fig10}
\end{figure*}

Among 37 HAEs used in this paper, six objects show flux excesses 
in NB428 of which only two have significant \lya\ emission lines.
The remaining 31 HAEs show no flux excesses in the NB428 image. 
Such a low fraction of LAEs well agrees with the past 
dual \lya\ and \ha\ emitter surveys 
\citep{Hayes:2010,Oteo:2015,Matthee:2016}. Their \lya\ lines are 
either absent or in absorption (LAAs). In fact, such a high fraction 
of LAAs is expected from the past deep spectroscopic surveys 
\citep{Shapley:2003,Reddy:2008,Hathi:2016}. \citet{Hathi:2016} 
reported that about a half of UV luminous galaxies (i$<$25 mag) have 
EW$_\mathrm{Ly\alpha}$$<$0 \AA\ and their median stellar mass is 
log(M$_\star$/\msun)=9.65$_{-0.02}^{+0.03}$. On the other hand, the 
median stellar mass of the HAEs is 10.12$_{-0.59}^{+0.50}$, which 
suggests that the high fraction of LAAs in HAEs is probably due to 
older stellar populations in our HAE samples.
In this section, we aim to interpret the origin of the \lya\ absorption.
The uniqueness of this study is that our samples are selected so that 
they are limited by \ha\ line flux based on the narrow-band imaging at 
NIR, and hence independent of \lya\ line flux. They are widely spread in 
UV luminosities. Although the \lya\ narrow-band imaging cannot provide 
the spectrum of the \lya\ line directly, we can put constraints on 
EW$_\mathrm{Ly\alpha}$ for individual galaxies even for UV faint objects 
(B$>25$ mag). Such measurements would be almost impossible with optical 
spectroscopy of a realistic observation time. 

HAEs are expected to be located at $z$=2.50--2.55 whose 
\lya\ emission can be covered by the narrow-band filter NB428 
(Fig.~\ref{fig1}). One caution is that the NB2315 filter suffers from
a small wavelength shift depending on the location of 
the detector as reported by \citet{Tanaka:2011}. The filter response 
curve shifts toward shorter wavelength up to $\sim$90 \AA\ at the 
edge of the detectors, which may affect filter coverage of \lya\ 
line of HAEs at $z=2.5$. However, we think this effect is small,
since \citet{Shimakawa:2014} have identified that the HAEs discovered by 
the NB2315 filter in another field are confirmed to fall within the 
redshift interval of 2.50--2.54 by a follow-up spectroscopic survey.  
Therefore, the NB428 filter should neatly cover the wavelength range
for \lya\ lines of our HAEs, and we do not take this effect into 
account in the following. 

We investigate what physical properties such as SFRs, specific SFRs, 
and EW$_\mathrm{H\alpha}$ are related to EW$_\mathrm{Ly\alpha}$ 
of HAEs. However, no clear relationship can be identified. In fact, 
the depth of the current narrow-band data is not enough to detect
\lya\ absorption and put a constrain on EW$_\mathrm{Ly\alpha}$ if it is 
smaller than $\sim-15$ \AA. Unfortunately, this limitation does not 
allow us to study the physical meanings of the \lya\ absorption features 
in most of the HAEs, and we require much deeper narrow-band imaging for 
this purpose. However, we can compare our HAEs and LAEs in 
Fig.~\ref{fig10} where EW$_\mathrm{Ly\alpha}$ is compared with (a) 
stellar masses on the left, and (b) dust reddening of UV continua on the 
right. Stellar masses and E(B$-$V) are derived from the SED-fitting and 
the UV slope as explained in \S2.5. Moderate correlations can be seen in 
these plots.

We should note that narrow-band selected LAEs tend to be biased towards 
high EW$_\mathrm{Ly\alpha}$. However, these figures can easily tell us 
the differences between \ha\ and \lya-flux limited samples; HAEs tend 
to be more massive and dustier than LAEs. Our HAEs seem to trace the 
EW$_\mathrm{Ly\alpha}$--E(B$-$V) relation found from the composite 
spectra of Lyman break galaxies at $z\sim$3 (\citealt{Shapley:2003}, 
see also \citealt{Cassata:2015}). 
This trend is also inferred from the anti-correlation 
between E(B$-$V) and the \lya\ photon escape fraction 
\citep{Hayes:2010,Matthee:2016}. 
We here note that a discrepancy between our LAE samples and LBGs 
\citep{Shapley:2003} at positive EW$_\mathrm{Ly\alpha}$ should be due 
to a sample selection effect. The weak anti-correlation between E(B$-$V) 
and EW$_\mathrm{Ly\alpha}$ can be interpreted as higher \hi\ covering 
fraction for galaxies with higher dust extinction as suggested by 
\citet{Reddy:2016b}. However, this scenario cannot explain some outliers 
seen in the upper right corner of Fig.~\ref{fig10}b. They show the reddest UV 
slopes, E(B$-$V) $>0.25$, and their locations are significantly deviated from
the distributions of HAEs and LAEs on this diagram. Such a reverse trend has
also been reported by \citet{Matthee:2016}, which find a bimodal relation in the 
UV slope versus the \lya\ photon escape fraction of HAEs at $z=2.2$. 
Those dusty LAEs would need another physical mechanism to emit strong 
\lya\ line emission such as galactic outflows powered by
starbursts and/or AGNs 
\citep{Mas-Hesse:2003,Hashimoto:2013,McLinden:2014,Hayes:2015,U:2015}. 
To confirm this hypothesis, follow-up spectroscopy is required.

\section{Discussion}

We have traced LAEs at $z=2.53$ down to the stellar mass of 10$^{8}$ 
\msun\ and SFR of 0.2 \msun/yr by deep narrow-band imaging as well 
as using publicly available various large surveys data (e.g. SXDS, 
CANDELS, 3D-HST). With this unique data-set, we investigate the 
yet-unexplored lower mass regime of the two major scaling relations at 
high redshifts; i.e., mass--SFR and mass--size relations. The results 
presented in the previous section suggest that the LAEs show no 
especially high or low star-forming activities for their stellar masses. 
Also their stellar sizes are neither particularly large nor particularly 
small for their masses when compared to the normal star-forming 
galaxies. These similarities are consistent at least in the low-mass 
regime, while massive LAEs seem to show distinctive physical properties 
at the same time. This section provides some further discussion on the 
results, including some concerns about the current analyses.

\subsection{Similarities of physical properties in low-mass LAEs}

This work has compared the physical properties of LAEs with those of 
HAEs at the same redshift, 2.5. Our results suggest that LAEs are less 
massive, less active, less dusty and younger galaxies than HAEs, and 
are smaller than more massive star-forming galaxies. However, for a 
given fixed stellar mass, they show no special characteristics in terms 
of SFR or size compared to other normal star-forming galaxies. This 
means that the LAE selection just picks out low-mass star-forming 
galaxies preferentially. Our selection is based on \lya\ luminosity, 
\lya\ equivalent width, and F160W-band magnitude. It can trace 
star-forming galaxies with stellar masses down to $\sim$10$^{8.5}$ 
\msun, if they are located on the linearly extrapolated main sequence 
of higher mass galaxies to the lower mass regime 
\citep{Whitaker:2014b,Shivaei:2015,Tomczak:2016} (see \S2.5 and \S3.1). 

\begin{figure*}
	\includegraphics[width=130mm]{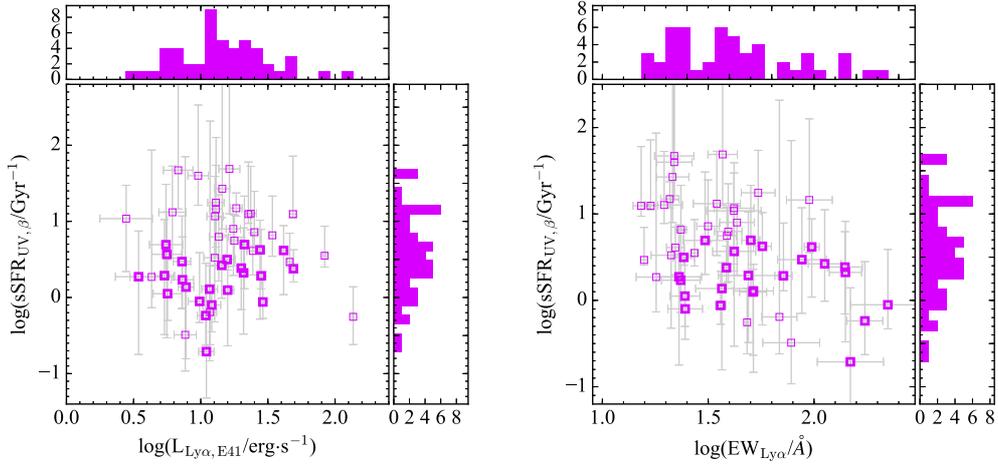}
    \caption{(a) The left panel shows \lya\ luminosity versus specific 
    SFR derived from UV luminosity. (b) The right panel represents 
    EW$_\mathrm{Ly\alpha}$ versus specific SFR of LAEs at $z=2.5$. The 
    SFRs are dust-corrected based on E(B$-$V) from the UV slope. The 
    symbols are the same as in Fig.~\ref{fig8}. Thick symbols indicate 
    LAEs at M$_\star\leq$10$^{10}$ \msun\ below the main sequence by 
    \citet{Shivaei:2015} shown in the Fig.~\ref{fig8}b.} 
    \label{fig11}
\end{figure*}

Figure~\ref{fig11} shows \lya\ luminosity and EW versus specific SFR 
(sSFR) of the LAEs. Thick symbols show less massive LAEs located below 
the main sequence of \citet{Shivaei:2015}. At the first sight, there are 
no strong trends in these diagrams, which suggests that enhanced 
star-forming activity is not required for the \lya\ photons to escape. 
Also, this indicates that our result is independent of the thresholds in 
\lya\ luminosity or EW. Therefore, star-forming activities may not be 
directly related to the escape mechanism of the \lya\ photons.
There is actually a weak correlation between EW$_\mathrm{Ly\alpha}$ and 
sSFR (Spearmans's correlation coefficient of $-$0.39). This is probably 
due to the increased UV continua in high sSFR galaxies, and/or this 
would be partly caused by the observational limitation. As mentioned in 
\S3.1, the narrow-band selection is insensitive to less active LAEs with 
low EW$_\mathrm{Ly\alpha}$, which depend on the depth of narrow-band 
and corresponding broad-band (B-band) images and the observed UV slopes 
of galaxies. Taking into account past studies (e.g. 
\citealt{Yajima:2012,Shibuya:2014b,Reddy:2016b}), in principle the \hi\ 
gas covering fraction and the dust obscuration are more likely to be 
controlling the \lya\ photon escape. 

While the results suggest that the activity of star-formation does not 
play a critical role in the \lya\ escape, we still have a concern about 
interpreting the similarities in star-forming activities of LAEs on the 
mass--SFR diagram. The results of this work are based on the SFR 
calibration from UV luminosities and continuum slopes of LAEs, 
following the \citet{Kennicutt:1998,Meurer:1999} prescriptions. This 
has a large uncertainty for young starburst populations, however, since
the SFR estimations based on their calibrations are no longer valid as 
cautioned in their studies. The \citet{Kennicutt:1998} calibration 
assumes a continuous, constant star-formation rate over a timescale of 
10$^8$ yrs or longer. They also mention that this conversion tends to 
underestimate SFRs for very young star-forming galaxies. 
\citet{Wuyts:2013} have reported a time dependence of the intrinsic 
\ha/UV flux ratio for various star-formation histories (see Fig.\ 3 in 
their paper). For instance, they show that the intrinsic \ha/UV ratio 
increases towards younger ages ($<$100 Myr) if a constant 
star-formation rate is assumed. \citep{Cervino:2016} have reported in 
detail that the ratio of ionizing and non-ionizing photons strongly 
depends on the constituting stellar populations of different ages 
\citep{Cervino:2016} determined by the star-formation histories. 
Therefore, even though galaxies have the same dust-corrected sSFR 
inferred from UV luminosities, they do not necessarily have the same 
amount of ionizing photons or \ha\ (\lya) fluxes between the two cases. 
However, we cannot accurately derive \ha\ emission flux of LAEs with 
the NB2315 filter data alone, since the transmission curve of NB428 
covers a wider redshift range than NB2315. 

In short, this work suggests that \lya\ luminosities 
and equivalent widths do not necessarily correspond to star-forming 
activities at least in young low-mass galaxies ($<$10$^{10}$ \msun), 
which is consistent with \citet{Hagen:2016}. Whereas, 
we should obtain more robust SFRs of LAEs independent of their 
star-formation histories, since those might have higher ionizing 
photons for a given sSFR inferred from UV luminosities. This factor 
is more important for SFR measurements of low-mass LAEs. There are 
actually low-mass star-forming galaxies at $z$$\sim$2 that have 
higher SFRs if \ha\ is used, than those measured from UV 
\citep{Shivaei:2015}. Indeed, there is a possibility that \lya\ photons 
can more easily escape from less massive systems. In addition to that, 
this work does not use low-mass star-forming galaxies, which do not 
show \lya\ in emission for comparison. A direct comparison of the 
properties of LAEs and non-LAEs at the same low-mass regime would be 
required to understand the mechanism of \lya\ escape in detail. 
Therefore, the current results are at this stage too early to conclude 
that LAEs are in the phase of secular evolution as is typical for 
normal star-forming galaxies in this mass range. A deep \ha\ line 
imaging survey (e.g. \citealt{Hayashi:2016}) enables us to trace 
low-mass star-forming systems that have been explored only by \lya\ 
emission so far, and to resolve such an issue. Last, our results are 
limited to EW$_\mathrm{Ly\alpha}>15$ \AA. The most serious bias in this 
work could be that the narrow-band imaging would miss the LAEs with low 
EW$_\mathrm{Ly\alpha}$. Thus, our sample may be missing UV luminous 
LAEs with low \lya\ luminosities like Lyman break galaxies. However, 
we stress that such a bias, if it presents, would rather strengthen 
our conclusion that high sSFR is not required for \lya\ photons to 
escape.

\subsection{Surface mass densities at z=2.5}

\begin{figure*}
	\includegraphics[width=170mm]{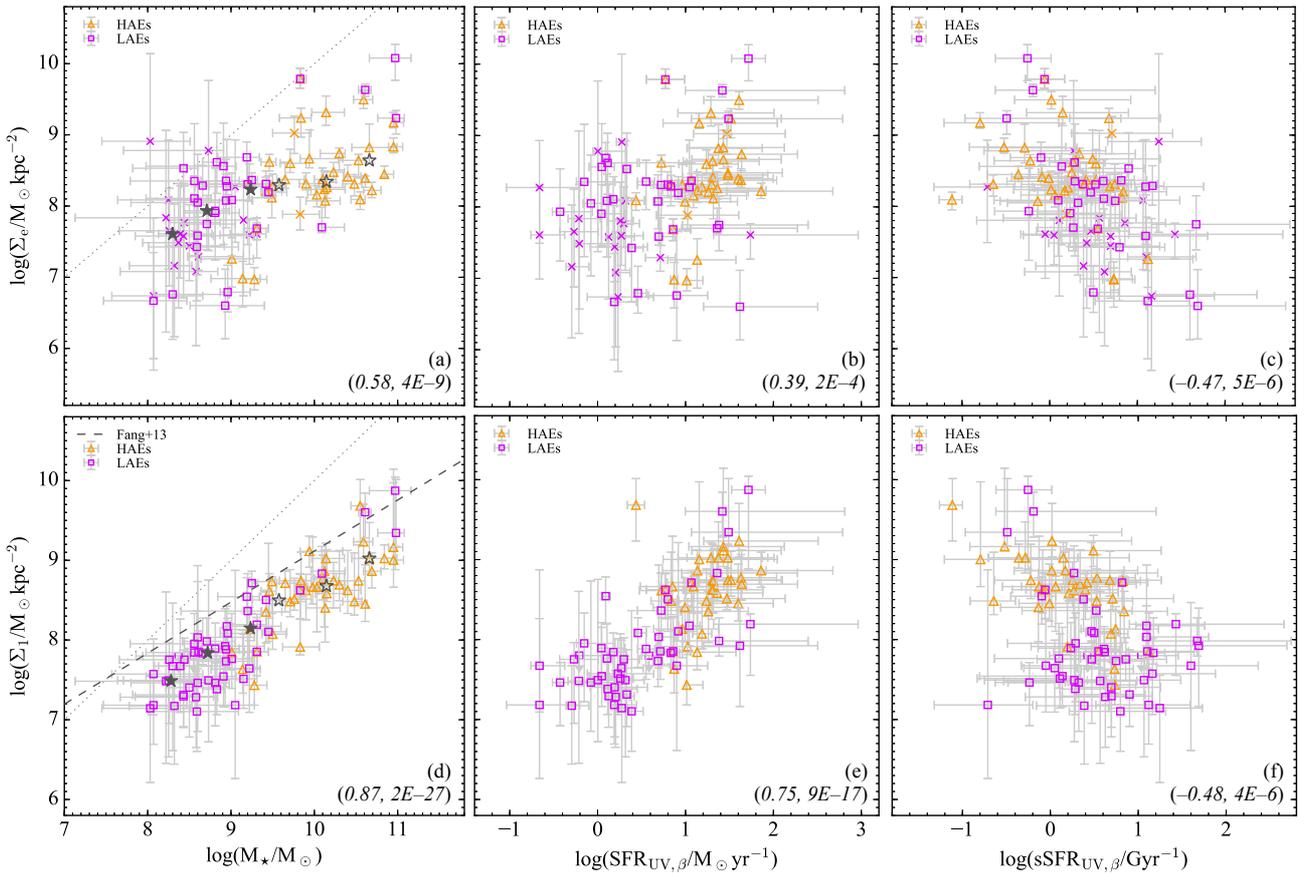}
    \caption{(a--c: from left to right in the upper part) Log stellar 
    mass, dust-corrected SFR, and dust-corrected sSFR versus log 
    effective mass surface density, (d--f: from left to right in the 
    lower part) log stellar mass, dust-corrected SFR, and 
    dust-corrected sSFR vs. log mass surface density within 1 kpc 
    radius. The symbols are the same as shown in Fig.~\ref{fig9}. 
    Dotted lines on the left panels indicate log(M$_\star$/\msun)= 
    log($\Sigma_\mathrm{e}$/\msun kpc$^{-2}$) (a) and 
    log($\Sigma_\mathrm{1}$/\msun kpc$^{-2}$) (d) for each diagram. 
    The dashed line indicates the mass--$\Sigma_\mathrm{1}$ relation 
    of green valley and red sequence galaxies at $z$$\sim$0, based on 
    the GALEX-SDSS cross-matched sample \citep{Fang:2013}. The 
    Spearman's rank correlation coefficients ($r$) and $p$ values are 
    inset on lower left side in each diagram in the form of ($r$, $p$).} 
    \label{fig12}
\end{figure*}

We investigate structure evolution of star-forming galaxies at $z$=2.5 
based on the LAE and HAE samples. Figure~\ref{fig12}a,c represent the 
surface density of stellar mass within effective radii 
($\Sigma_\mathrm{e}$=M$_\star$/2$\pi$R$_\mathrm{e}^2$) and that within
1 kpc radii ($\Sigma_\mathrm{1}$) as a function of stellar mass. Because 
of the large errors on effective radii, $\Sigma_\mathrm{e}$ have large 
uncertainties in the low stellar mass range. $\Sigma_\mathrm{1}$ are 
derived from the SED fitting with {\sc fast} \citep{Kriek:2009} based on 
the HST band photometries (F125,160W and F606,814,140W if available) 
within a fixed aperture radius of 1 kpc on a physical scale (0.3 arc sec
diameter). Thus the $\Sigma_\mathrm{1}$ is computed from  
small aperture measurements, which stellar masses (M$_\star$) are 
obtained for the whole galaxy. In the SED fitting, we set the 
same initial parameters as described in \S2.5 assuming the delayed 
exponentially declining star-formation history.

If we limit our sample only to HAEs, we do not see any strong 
correlation between stellar mass and $\Sigma_\mathrm{e}$, while low-mass 
LAEs ($<10^{10}$ \msun) show a moderate positive relation between them 
(Spearman's rank correlation coefficient is 0.37 corresponding to greater 
than 95\% confidence level). On the other hand, we see an even tighter 
correlation between stellar mass and $\Sigma_\mathrm{1}$. The 
coefficient is 0.87 with more than 99.9 \%\ confidence level for the 
combined sample of HAEs and LAEs. The distribution of galaxies of our 
sample on this diagram is consistent with other works such as 
\citet{Fang:2013,Tacchella:2015} for the local and $z$$\sim$2 star-forming 
galaxies, respectively. The positive correlations seen in these two 
diagrams offer the strong evidence for a vigorous mass growth in the early 
phase of galaxy formation. For massive galaxies above 
log(M$_\star$/\msun)$\gtrsim$9.5, stellar mass -- $\Sigma_\mathrm{e}$ 
relations are flattened and/or scattered. This suggests that the central 
part of galaxies is quickly built up and the inner mass density reaches to 
nearly a constant value as a function of mass (time). This indicates the 
inside-out growth of galaxy formation 
\citep{Nelson:2012,Nelson:2013,Nelson:2016a}. 

We confirm the existence of moderate or strong correlations of SFR and 
sSFR with mass surface densities 
($\Sigma_\mathrm{e}$ and $\Sigma_\mathrm{1}$) 
(Fig.~\ref{fig12}b,c,e,f) with the Spearman's rank correlation 
coefficient of $r=0.39$ in $\Sigma_\mathrm{e}$--SFR, $r=-0.47$ in 
$\Sigma_\mathrm{e}$--sSFR, $r=0.75$ in $\Sigma_\mathrm{1}$--SFR, and 
$r=-0.48$ in $\Sigma_\mathrm{1}$--sSFR, respectively. The corresponding
significance levels are more than 99.9\%. However, such 
correlations disappear except for the $\Sigma_\mathrm{1}$--SFR relation, 
if we limit our sample to those with low surface densities $\lesssim$10$^{8}$ 
\msun/kpc$^2$ (see also \citealt{Franx:2008,Whitaker:2016}). As the 
effective mass surface densities grow, star-forming activities 
characterised by the specific SFR tend to decline, while they tend to be more
spread in SFRs (Fig.~\ref{fig12}b,c). On the other hand, mass surface 
densities within 1 kpc strongly correlate with the total stellar masses and 
SFRs, and yet there is still a moderate anti-correlation between 
$\Sigma_\mathrm{1}$ and sSFR. Such a surface density effect is thought 
to be one of the major factors leading to a quenching of star formation 
activities (\citealt{Franx:2008,Fang:2013,Omand:2014,Woo:2015,Whitaker:2016}, 
but see \citealt{Lilly:2016}). This may be qualitatively consistent with the 
morphological/gravitational quenching 
\citep{Martig:2009,Davis:2014,Genzel:2014}. 

In section \S3.1 and \S3.2, we not only confirm the similarities of 
star-forming activities and size distributions of LAEs at the 
low-mass regime, but we now also find a possible uniqueness of massive 
LAEs which tend to be compact and less active for a given stellar mass. 
Fig.~\ref{fig12} exactly provides us with the combined information of 
these two results. Massive LAEs are located at the edge on the sSFR 
versus mass surface density diagrams in the meaning that they tend to 
be compact and less active. This trend becomes more prominent when we 
use SFR$_\mathrm{UV,sed}$ (with dust correction based on the 
SED-fitting) and SFR$_\mathrm{SED}$ (SED-inferred SFR). These less 
active massive LAEs would be post-starburst-like galaxies as discovered 
by \citet{Taniguchi:2015}, which might have experienced a nuclear 
starburst due to such major mergers and a subsequent quenching. 
Combined with a clear \lya\ escape seen in them, a favourable 
interpretation of such massive LAEs is that a nuclear starburst invokes 
a strong galactic outflow, which clears surrounding gas along the line 
of sight, and eventually allows \lya\ photons to escape. So far, we do 
not have a strong evidence for this hypothesis. Note that they do not 
have strong AGNs as they are not identified so by X-ray 
\citep{Ueda:2008} or MIPS 24$\mu$m \citep{Dunlop:2007} data.

In Table~\ref{tab3}, we summarise the median values and 1 sigma 
scatters of dust-uncorrected SFRs (SFR$_\mathrm{UV,obs}$), SFRs 
corrected for dust extinctions in several different ways 
(SFR$_\mathrm{UV,\beta}$, SFR$_\mathrm{UV,sed}$, and SFR$_\mathrm{SED}$),
effective radii (R$_\mathrm{e}$), and mass surface densities within 
effective radii ($\Sigma_\mathrm{e}$) and those within 1 kpc radii 
($\Sigma_\mathrm{1}$) for the LAEs in three different stellar mass bins 
($<10^{8.5}$, $10^{8.5}$--$10^{9}$, and $10^{9}$--$10^{10}$ \msun).

\begin{table}
    \centering
	\caption{Median values and 1 sigma scatters of log SFR$_\mathrm{UV,obs}$, 
    SFR$_\mathrm{UV,\beta}$, SFR$_\mathrm{UV,sed}$, SFR$_\mathrm{SED}$, 
    R$_\mathrm{e}$, $\Sigma_\mathrm{e}$, $\Sigma_\mathrm{1}$ of LAEs 
    at $z=2.5$ in three different stellar mass bins ($<10^{8.5}$, 
    $10^{8.5}$--$10^{9}$, and $10^{9}$--$10^{10}$ \msun).}
 	\label{tab3}
	\begin{tabular}{lrrr}
		\hline
		log Stellar mass            & 8.30$_{-0.16}^{+0.11}$ & 8.71$_{-0.14}^{+0.18}$ & 9.24$_{-0.16}^{+0.25}$ \\
		\hline
        log SFR$_\mathrm{UV,obs}$   & $-$0.05$_{-0.33}^{+0.24}$ &  0.20$_{-0.31}^{+0.30}$ & 0.30$_{-0.39}^{+0.45}$ \\ 
        log SFR$_\mathrm{UV,\beta}$ & 0.19$_{-0.49}^{+0.27}$  &  0.39$_{-0.45}^{+0.54}$ & 0.75$_{-0.75}^{+0.41}$ \\ 
        log SFR$_\mathrm{UV,sed}$   & 0.09$_{-0.48}^{+0.69}$  & 0.46$_{-0.51}^{+0.68}$ & 0.70$_{-0.59}^{+0.53}$ \\ 
        log SFR$_\mathrm{SED}$      & $-$0.27$_{-1.31}^{+1.59}$   & 0.04$_{-1.50}^{+1.39}$ & 0.56$_{-0.70}^{+0.52}$ \\ 
        log R$_\mathrm{e}$          & $-$0.01$_{-0.37}^{+0.27}$   & 0.04$_{-0.44}^{+0.64}$ & 0.11$_{-0.22}^{+0.26}$ \\ 
        log $\Sigma_\mathrm{e}$     & 7.61$_{-0.66}^{+0.64}$   & 7.93$_{-1.38}^{+0.82}$ & 8.23$_{-0.58}^{+0.57}$ \\ 
		\hline
		log Stellar mass            & 8.28$_{-0.15}^{+0.13}$ & 8.72$_{-0.14}^{+0.18}$ & 9.24$_{-0.16}^{+0.25}$ \\
		\hline
        log $\Sigma_\mathrm{1}$     & 7.45$_{-0.25}^{+0.19}$   & 7.84$_{-0.41}^{+0.17}$ & 8.15$_{-0.54}^{+0.41}$ \\ 
        \hline
    \end{tabular}
\end{table}

\subsection{Are the LAEs building blocks of local L* galaxies?}

\begin{figure}
	\includegraphics[width=80mm]{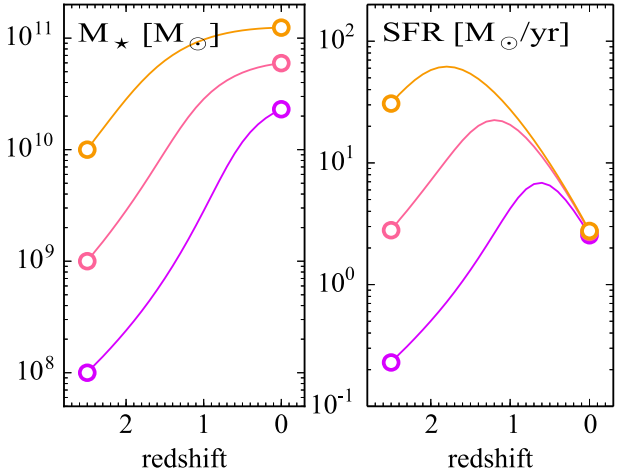}
    \caption{The predicted evolution in stellar mass (a; left) and SFR 
    (b; right) of low-mass galaxies on the main sequence from $z$=2.5 to 0. 
    Purple (top), pink (middle), and orange (bottom) curves indicate the 
    evolutionary track from the initial masses of $1\times10^{8}$, 
    $10^{9}$, and $10^{10}$ \msun, respectively. Those growth histories 
    are based on the assumption that the galaxies grow along the main 
    sequence at each redshift. The main sequences as a function of 
    redshift are derived from the prescription by \citet{Tomczak:2016}.}
    \label{fig13}
\end{figure}

Another interesting question is whether the LAEs evolve into typical L* 
galaxies in the local Universe. Past clustering analyses suggest that 
LAEs at $z\sim2$ can be the building blocks of local L* galaxies like 
the Milky Way \citep{Gawiser:2007,Guaita:2010,Guaita:2011}. We therefore 
perform an experiment to discuss how the mass of these low-mass LAEs 
grows in time (redshift) based on a simple assumption that they are 
forced to grow along the redshift-dependent main sequence. A similar 
test has been performed by \citet{Tomczak:2016}, but in this paper, we 
discuss much lower mass galaxies at high redshift than those in 
\citet{Tomczak:2016}.

Note that such an analytical simulation gives us only a rough estimate 
and the actual evolution would depend on galaxy mergers and/or 
quenching of star formation. In a merger case, the estimated masses 
would be lower limits, as mergers can grow their masses further. In a 
quenching case, some LAEs may well be quenched by now and would have 
been dropped from the main sequence. Furthermore, some galaxies would 
appear suddenly on the main sequence either from above or below the 
main sequence at their birth time. We do not take into account all 
these complicated processes, and thus the results presented here 
should be received with caution.

Nevertheless, under those assumptions, the stellar mass as a function 
of redshift can be presented by the following equation;

\begin{equation}
    m_{t}(z)=m_0 + \Sigma_{i=0}^{z}[(1-R(t_{z-z_i})) \times \psi(z_i,m_i) \times \delta t_{z-z_i}].
	\label{eq2}
\end{equation}

We set the initial redshift of $z_0$=2.5 and the time interval of 
$\Delta z=-0.1$. $m_0$ is the initial mass at $z=z_0=2.5$. 
$\delta t_{z-z_i}$ [yr] is the age interval from the redshift $z_i$ to 
$z$ when we have the total stellar mass of $m_t(z)$. $\psi(z_i,m_i)$ 
is SFR at the given stellar mass of $m_i$ and the redshift of $z_i$, 
assuming the redshift-dependent main sequence reported by ZFOURGE 
survey \citep{Tomczak:2016}. We must caution that their prescription 
does not cover low-mass galaxies at $z>2$. However, we assume this 
model works well even for our lower mass LAEs since they follow the 
extrapolated main sequence line as shown in Fig.~\ref{fig8}. The 
equation is the following (eq.~4 in their paper);

\begin{eqnarray*}
    \log[\psi(z_i,m_i)] &=& s_0(z_i) -\log[1+(m_i/M_0(z_i))^{-\gamma}] \\
    s_0(z_i) &=& 0.448 +1.220 z_i -0.174 z_i^2 \\
    \log[M_0(z_i)] &=& 9.458 +0.865 z_i -0.132 z_i^2 \\
    \gamma &=& 1.091.
	\label{eq3}
\end{eqnarray*}

$R(z_i)$ indicates the mass return rate from stars to interstellar 
matter determined by the mass loss rate of massive stars (on a 
timescale less than $1\times10^8$ yr), AGB stars, and type-Ia 
supernovae (the latter two are dominant in later phases). The mass 
return rate strongly depends on the assumed stellar initial mass 
function and it is also slightly dependent on the used model. For 
example, \citet{Moster:2013} employs the mass return rate predicted 
by the stellar population synthesis model of \citet{Bruzual:2003} 
with \citet{Chabrier:2003} IMF. This provides the mass return rate 
of 12\% higher than that used in \citet{Segers:2016} which is based 
on a different simple stellar population model of 
\citet{Wiersma:2009}. We use the former prescription, which gives 
the following equation (eq.~16 in \citealt{Moster:2013});

\begin{equation*}
    R(t_{z_-z_i})=0.05 \ln[(t_{z-z_i}+3\times10^5)/3\times10^5].
	\label{eq4}
\end{equation*}

The result of this simple analytical simulation shows that LAEs 
found in this observation can grow to galaxies with several times
$10^{10}$ \msun\ (Fig.~\ref{fig12}a). Figure~\ref{fig12}b shows 
their SFR histories. This suggests that star-forming activities in 
low-mass galaxies at $m_0=1\times10^8$ \msun\ are modest and cannot 
exceed the SFR of 10 \msun/yr during their growth histories. The 
predicted main sequence has a nearly flat slope above 
$1\times10^{10}$ \msun\ at $z$=0, and SFRs are all similar 
irrespective of initial stellar masses.

As a result, this simulation shows that the Milky Way galaxy with a 
stellar mass of $6.4\pm0.6\times10^{10}$ \msun\ \citep{McMillan:2011} 
can be a descendant of one of those LAEs with a stellar mass of 
$\sim1\times10^{8.5}$ \msun\ at $z=2.5$. Some fraction of the 
low-mass LAEs at $z=2.5$ may evolve into local L* type galaxies like 
the Milky Way.
On the other hand, HAEs with 
M$_\star>10^{10}$ \msun\ seem to have a potential to grow into 
very massive objects with M$_\star>10^{11}$ \msun\ if they retain 
substantial gas reservoirs until later times.

\subsection{Ly$\alpha$ absorbers traced by narrow-band imaging}

\begin{figure}
	\centering
	\includegraphics[width=75mm]{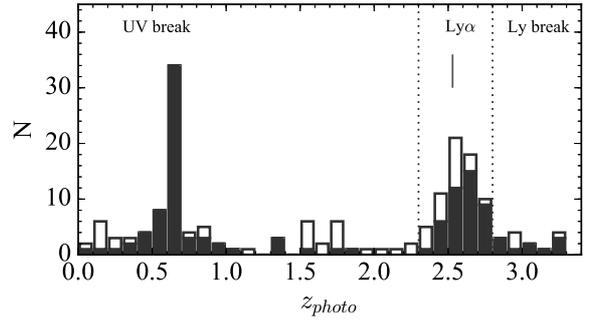}
    \caption{Photometric redshift distribution of the NB absorbers. 
    We employ the best photometric redshifts ($z_\mathrm{best}$) 
    in their catalogue. Filled and open histograms present the NB 
    emitters with F160W$\le$25 and F160W$>$25 mag, respectively. NB 
    absorbers located at $z_\mathrm{best}$=2.3--2.8 are selected as 
    \lya\ absorbers (LAAs).}
    \label{fig14}
\end{figure}

Interestingly, a part of our HAE samples show \lya\ absorption rather 
than emission as shown in \S3.3. The weak negative trend between 
E(B$-$V) and EW$_\mathrm{Ly\alpha}$ is consistent with the past deep 
spectroscopic studies of Lyman Break galaxies at $z>2$ 
\citep{Shapley:2003,Cassata:2013,Hathi:2016} and agrees with the 
anti-correlation between E(B$-$V) and the \lya\ photon escape 
fraction reported by the past dual \ha\ and \lya\ emitter 
surveys \citep{Hayes:2010,Matthee:2016} if we assume that the
\lya\ EW is roughly comparable to the \lya\ photon escape fraction. 

The fact that strong \lya\ absorption features in a part of HAEs 
indicates the potential use of narrow-band imaging to effectively 
search for \lya-absorbed galaxies in narrow redshift slices by 
identifying the narrow-band flux deficits. In fact, this method is 
already demonstrated by \citet{Hayashino:2004} in the familiar 
protocluster field, SSA22 at $z$=3.1 discovered by 
\citep{Steidel:1998}. Here, we evaluate the availability of this 
technique again by incorporating the photometric redshift catalogue 
by 3D-HST \citep{Momcheva:2016} in the UDS-CANDELS field. 

We first search for NB absorbers showing $3\sigma$ deficits in NB428 
flux relative to B-band. We set the quite similar selection criteria 
as used for the identification of LAEs in this paper; (1) the colour 
cut of NB$-$B$>$0.476 mag corresponding EW$_\mathrm{Ly\alpha}=-9$ 
\AA\ in the rest frame at $z=2.53$, and the limiting magnitudes of 
$5\sigma$ in B band (27.89). For NB absorbers without narrow-band 
detections, 2$\sigma$ lower limit of NB magnitude are assumed (see 
also Appendix B). 

Figure~\ref{fig14} presents photometric redshift distribution of 
obtained NB absorbers. The histogram suggests that this technique 
allows us to find \lya\ absorbers (LAAs) at $z$=2.5 quite 
effectively. The number of LAAs at photo-$z$ =2.3--2.8 is 65 is 
slightly larger than that of our LAE sample (50). The number ratio 
of LAAs to LAEs is roughly consistent with the result by recent deep 
spectroscopic survey \citep{Hathi:2016}. Another spike seen at 
$z$$\sim$0.6 is due to UV break (B2640). A small difference of 
central wavelength between NB428 and B (170\AA) allows us to detect 
this break feature (see Appendix B). The UV break around 2640 \AA\ 
in the rest frame is known to be apparent in passive evolving 
galaxies (e.g. \citealt{Cimatti:2008}). Indeed, we roughly check and 
confirm those red colours and spheroidal morphologies based on HST 
ACS/WFC3 images by visual inspection.

\begin{figure}
	\centering
	\includegraphics[width=75mm]{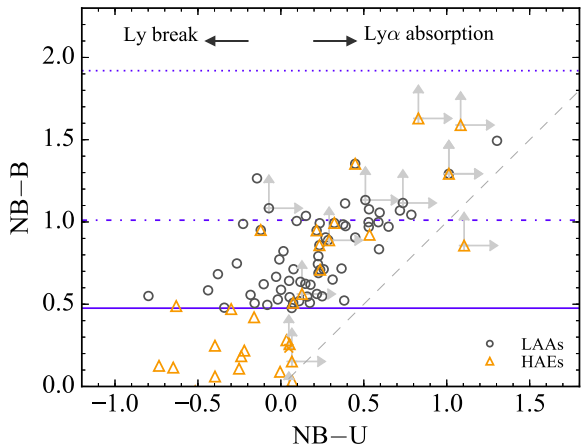}
    \caption{Colour-colour diagram between NB428$-$U and NB428$-$B. 
    Black circles and orange triangles are \lya\ absorbers (LAAs) 
    and HAEs, respectively (some of them overlap). Cross symbols show 
    HAEs that cannot be seen in either NB428 and U bands, and thus 
    those NB$-$U colours cannot be constrained at all. The blue solid 
    horizontal line corresponds to the EW threshold of $\sim-$9 \AA\ 
    in the rest frame. The dash-dot and the dotted blue lines 
    represent EW$_\mathrm{Ly\alpha}$=$-$15 and $-$20 \AA, 
    respectively. The grey dashed line indicates NB$-$U=NB$-$B.}
    \label{fig15}
\end{figure}

We stress that the narrow-band deficits seen in LAAs should not be 
caused by Lyman Break feature at $\lambda$$<$1216 \AA. 
Figure~\ref{fig15} show a colour-colour magnitude diagram of NB$-$U 
and NB$-$B. U-band filter covers the rest-frame wavelength of 
955--1164 \AA\ at $z$=2.53 which samples \lya\ forest continua. Thus 
U-band magnitude is basically fainter than B-band and narrow-band 
magnitudes due to absorption by intergalactic medium (IGM) in 
addition to dust reddening. For example, typical IGM transmission at 
$z$$\sim$2 is about 80\% based on the calculation using BOSS \lya\ 
forest catalogue \citep{Lee:2013}, which corresponds to absorption of 
$-$0.24 mag in U-band. Nevertheless, the figure shows that LAAs with 
larger \lya\ absorption EW (higher NB$-$B) tend to be redder even in 
NB$-$U colour, which indicates that the NB deficits in LAAs should be 
caused by \lya\ absorption. In particular, the HAE samples at $z$=2.5 
selected independently by narrow-band observations at NIR also 
overlap with that of LAAs on this diagram. This strongly supports the
fact that the NB deficits are caused by \lya\ absorptions rather than 
the Lyman break feature. 

Such a strong \lya\ absorption is though to be due to high covering 
fraction (but not a unity) of \hi\ clouds with high column densities 
N(\hi) $>10^{21}$ cm$^{-2}$. \citet{Reddy:2016a,Reddy:2016b} have 
shown the composite FUV spectrum of Lyman Break galaxies at $z\sim3$, 
showing a blue-shifted \lya\ absorption feature 
due to the outflowing dense neutral hydrogen gas in the line of sight
similar to what we see in some dwarf galaxies in the local Universe
\citep{Kunth:1994,Atek:2009a,Rivera:2015}. Another causal 
factor would be stellar \lya\ absorption, however, this may be 
cancelled out by \lya\ emission at least for star-forming galaxies 
according to the theoretical model by \citet{Pena:2013}. Given the higher 
dust extinction of non-LAEs than that of LAEs 
\citep{Hathi:2016}, the former case (higher \hi\ covering fraction)
seems to be the dominant factor, since the dust reddening is correlated to
the \hi\ covering fraction \citep{Reddy:2016b}. We should note that our 
narrowband method could considerably underestimate \lya\ luminosity 
and EW$_\mathrm{Ly\alpha}$, depending on the locations of \lya\ emission 
or absorption with respect to the response curve of the narrowband filter. 
Complicated spectra at $\lambda\sim1216$ \AA\ including \lya\ emission 
and/or absorption prevent us from quantifying accurate \lya\ 
emissivities of galaxies only with the narrowband photometry. Therefore, 
we cannot rule out the possibility that these HAEs showing the 
narrowband deficits actually show \lya\ emission line as well. 

A problem is that our narrow-band technique allows us to securely search 
for LAAs only with EW$_\mathrm{Ly\alpha}<-9$ \AA, and we are not able to
detect the non-LAEs of course with the typical EW$_\mathrm{Ly\alpha}\sim0$ 
\citep{Cassata:2015,Hathi:2016}. Furthermore, ultra-deep narrow-band 
imaging is required to measure EW values of strong \lya\ absorbers by
reaching the bottom of the line.
Even if we conduct an observation with four times longer exposures (10 hrs),
we would be able to constrain EW$_\mathrm{Ly\alpha}$ 
only down to $-$20 \AA\ in the rest frame. Considering these factors, 
the available EW$_\mathrm{Ly\alpha}$ range with this method is actually 
quite limited. For example, our deep narrow-band observation can 
determine EW$_\mathrm{Ly\alpha}$ in the range between $-9$ and $-15$ 
\AA. This prevents us from studying the physical properties of LAAs over a
range of \lya\ absorption strength comprehensively.
This method, however, works well in searching for distant LAAs in a certain 
redshift slice, as demonstrated in this work.

\subsection{Supplementary comments}

Last, we complement the information about the LAE sample based on 
the narrow-band observations. However, note that it does not affect 
our main results or conclusions.

We also search for clumpy galaxies in the LAEs based on their WFC3/IR 
images. Our visual inspection finds only a few LAEs with multiple 
stellar components. This fraction is much smaller compared to the 
fraction of clumpy galaxies, $\sim$10--30\% seen in UV bright galaxies 
at $z\sim2$ \citep{Shibuya:2016}, or $\sim$40\% in \citet{Tadaki:2014}.
However, this difference may not be intrinsic, since we cannot resolve
the internal structures of LAEs due to their intrinsically small sizes
and their faintness.

To avoid a selection bias, this work does not exclude the mid-IR 
bright galaxy (LAE-31 in the Table~\ref{tab2}) detected in the Spitzer 
UKIDSS Ultra Deep Survey (SpUDS; \citealt{Dunlop:2007}). We also check 
the AGN population based on the catalogue by the Subaru/XMM-Newton 
Deep Survey \citep{Ueda:2008}. However, we find no X-ray counterparts 
within 3 arcsec radius of each of the LAEs. 

This work has presented that most of the HAEs do not show \lya\ line 
in emission. Rather, they sometimes show \lya\ line in absorption. On 
the flip side, \citet{Steidel:2011} still find a diffuse \lya\ halo in 
the composite \lya\ narrow-band image of such \lya-absorption systems. 
\citet{Matthee:2016} also find diffuse \lya\ haloes in 
HAEs based on the dual \ha\ and \lya\ emitter surveys at $z=2.23$. In 
order to examine the presence of \lya\ halos in our HAE samples, we stack 
the \lya\ images of HAEs based on {\sc imcombine} task of {\sc iraf} 
by median, and then estimate \lya\ luminosity with different aperture 
diameter of 1.5, 2.0, 4.0, 6.0, and 8.0 arcsec with the SExtractor, 
which correspond to 9.9, 14.6, 31.5, 47.8, and 64.0 physical kpc, 
respectively. The composite \lya\ line image seems to be extended, 
since we find that \lya\ emission line luminosity roughly increases as
L$_\mathrm{Ly\alpha}=-3.72+0.49\times$(aperture diameter in arcsec) 
E41 erg/s. However, this luminosity growth is not significant if we 
take account of the errorbars. Therefore, we could not confirm a clear 
detection of \lya\ haloes with our data.

\section{Conclusions}

We have conducted an optical narrow-band imaging survey with the 
Suprime-Cam to search for LAEs at $z=2.5$ in the UDS-CANDELS field 
where we have already identified many HAEs in the same redshift slice 
by our previous narrow-band imaging survey at NIR. Together with 
existing WFC3/IR data from the HST and multi-band photometries from 
various large programs, we identify 50 LAE candidates at $z$=2.53 down 
to stellar mass of 10$^8$ \msun and SFR of 0.2 \msun/yr. Our sample 
are limited by \lya\ luminosities ($>4.40\times10^{41}$ erg/s in 1.5 
arcsec aperture diameter), \lya\ equivalent widths ($>15$ \AA), and 
H$_\mathrm{F160W}$ band magnitude ($<26.89$ mag). We combine 37 HAEs 
from our previous work \citep{Tadaki:2013}, and investigate the 
physical properties of LAEs and HAEs such as their SFRs and sizes. 

We compare star-forming activities and stellar angular sizes of LAEs 
with those of HAEs and other star-forming galaxies in the literature. 
The F160W/IR images for galaxies at $z=2.53$ are not affected by \hb\ 
or \oiii\ lines, which thus provide us with the accurate stellar 
masses and the sizes of the LAEs and HAEs. The depth of our 
narrow-band and B-band data allows us to explore star-forming galaxies 
with lower stellar masses down to $\gtrsim$10$^8$ \msun.
As a result, we find that LAEs follow the extrapolated line of the 
mass--SFR relation (star-forming main sequence) to the lower-mass end.
The result is in good agreement with and complementary to 
\citet{Hagen:2016} which have reported no difference between LAEs and 
controlled sample for comparison on the mass--SFR diagram at the 
similar redshift. The \lya\ luminosities and EWs do not strongly 
correlate with sSFR, indicating that star-forming activities do not 
contribute directly to the \lya\ photon escape mechanism. As suggested 
by recent studies (e.g. 
\citealt{Yajima:2012,Shibuya:2014b,Reddy:2016b}), gas covering 
fraction and dust extinction would be the primary key factors. Indeed, 
we identify \lya\ absorption features in a part of the HAEs 
individually, which tend to be dustier and more massive systems. 

Since NB428 filter can cover \lya\ emissions of the existing HAEs 
discovered by the past narrow-band (NB2315) imaging at NIR, the 
combination of these two narrow-band filter allow us to investigate 
\lya\ emissivities of HAEs. However, we detect significant flux 
excesses in only two HAEs. Six out of 37 HAEs have
positive \lya\ EWs, of which 2 show significant \lya\ flux excesses 
at more than 3 sigma levels. Such a small fraction of LAEs among HAEs 
is consistent with the past similar studies 
\citep{Hayes:2010,Matthee:2016}, although it should be noted that the 
percentage should depend on the survey depths, photometric aperture size, 
physical properties of the parent HAE samples, and so on 
\citep{Matthee:2016}. For example, photometric aperture size of 1.5 
arcsec diameter used in this work is a half of that used 
in \citet{Matthee:2016}. We thus focus more on the 
\lya\ emissivity or its escape fraction along the line of sight. 

Instead, we find flux deficits in the narrow-band for a larger 
fraction of our HAE sample. The flux-limited LAEs and HAEs are quite 
distinctively distributed on the EW$_\mathrm{Ly\alpha}$ versus stellar 
mass or EW$_\mathrm{Ly\alpha}$ versus E(B$-$V) diagrams. We conclude 
that LAEs have similar general properties as those of normal 
star-forming galaxies. However, they also remain as a unique 
population (low-mass young galaxies) because the majority of HAEs are 
not LAEs. 

LAEs (low-mass star-forming galaxies) seem to share the same 
mass-size relation of massive star-forming galaxies within errors. 
On the other hand, four out of five massive LAEs at 
log(M$_\star$/\msun)$>$9.5 have compact structures and are located 
close to the mass-size relation of early-type galaxies rather than 
that of late-type galaxies reported by \citet{Wel:2014}. Moreover, 
they have low sSFRs and high mass surface densities. Such massive 
LAEs have experienced nuclear starbursts in the past, which clear 
the surrounding gas by a galactic wind, allowing the \lya\ photon 
to escape from the systems as detected in our narrow-band \lya\ 
imaging.

Finally, we demonstrate the unique narrow-band technique to search 
for \lya\ absorbers (LAAs), which are observed as showing flux 
deficits in the narrow-band instead of flux excesses. Whilst this 
method can effectively trace LAAs in a certain redshift slice, we 
should note that the sample is limited to a narrow range of 
EW$_\mathrm{Ly\alpha}$ in absorption due to observational 
limitations.

In order to generalise our results to normal star-forming galaxies,
we will need to investigate other types of galaxy populations than 
LAEs at the same redshift and in the same stellar-mass range, and 
compare their physical properties with those of LAEs. We are now 
conducting systematic ultra-deep narrow-band imaging survey of 
low-mass, low-luminosity HAEs (Kodama et al.) for this purpose.

\section*{Acknowledgements}

Grounded on the data collected at the Subaru Telescope, which is 
operated by the National Astronomical Observatory of Japan. This 
work is also based on observations taken by the 3D-HST Treasury 
Program (GO 12177 and 12328) with the NASA/ESA HST, which is 
operated by the Association of Universities for Research in 
Astronomy, Inc., under NASA contract NAS5-26555. This work is 
subsidized by JSPS KAKENHI Grant Number 15J04923. This work was 
also partially supported by the Research Fund for Students (2013) 
of the Department of Astronomical Science, SOKENDAI (the Graduate 
University for Advanced Studies). Data analysis was in part 
carried out on a common use data analysis computer system at the 
Astronomy Data Centre, ADC, of the National Astronomical 
Observatory of Japan. We thank the referee for helpful comments. 
R.S. thanks Dr. Masami Ouchi for useful discussions. R.S. and T.S. 
acknowledges the support from the Japan Society for the Promotion 
of Science (JSPS) through JSPS research fellowships for young 
scientists. Y.M. is supported by KAKENHI (No. 20647268). T.K. 
acknowledges KAKENHI (No. 21340045 and 24244015).







\bibliographystyle{mnras}
\bibliography{bibtex_library} 



\appendix
\section{Mass--SFR diagrams}

The choice of SFR calibration among UV+IR, UV+E(B$-$V) and SED-fitting 
at UV--optical can affect the resultant slope of the main sequence. In 
particular, since the most dust obscured star-forming regions tend to 
be missing in the UV images and sometimes even in the optical images, 
the UV slopes of dusty objects are actually largely determined by outer 
regions, which are relatively more transparent due to small amounts of 
dust. This would then lead to an underestimation of E(B$-$V). In fact, 
recent spatially resolved analyses suggest a strong radial gradient in 
dust extinction in massive star-forming galaxies \citep{Tadaki:2014b, 
Nelson:2016b,Tadaki:2016} in the sense that galaxy centres have 
much heavier dust obscuration. Moreover, \citet{Tadaki:2015} unveil the 
nuclear components hidden by dust in HAEs at $z>2$ by the spatially 
resolved dust continuum observations with Atacama Large 
Millimeter/submillimeter Array (ALMA). At the low-mass end, the surveys 
limited by ionizing photon flux (such as LAEs and HAEs) tend to miss 
post-starburst galaxies, which still show detectable UV luminosities. 
These factors would lead to the observed flatter main sequence based on 
\ha.

The \ha\ luminosities with dust corrections by the Balmer decrement 
(\ha\ and \hb\ ratio) provide the best, most accurate measurements of 
on-going SFR in the optical-to-NIR regime. This is largely independent 
of past star-formation histories, and relatively insulated from the 
influence of dust attenuation as compared to UV continuum. However, a 
part of the filter response curve of NB428 for \lya\ emission is not 
overlapped with that of NB2315 for \ha\ emission, unfortunately. This 
small mismatch prevents us from robustly estimating SFRs by \ha\ 
luminosities obtained from the narrow-band imaging data 
\citep{Tadaki:2013} for our LAEs. 

\begin{figure*}
	\includegraphics[width=175mm]{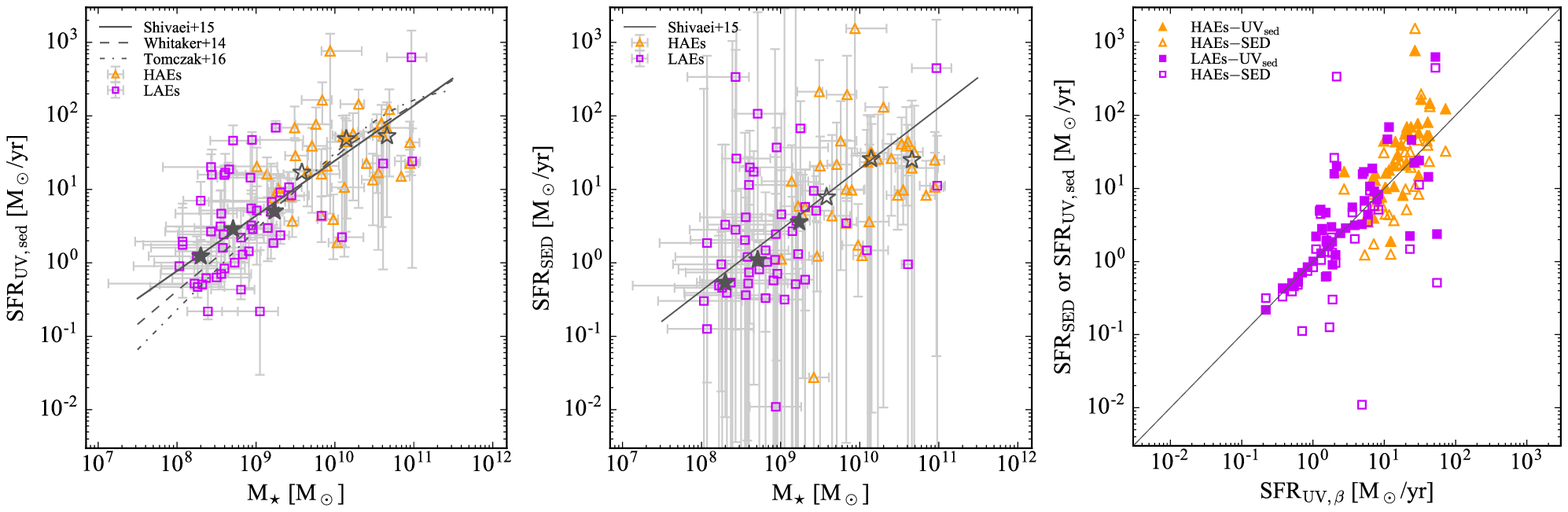}
    \caption{The stellar mass versus dust-corrected SFR, 
    SFR$_\mathrm{UV,sed}$ where dust reddening is obtained from the 
    SED-fitting (a: left) and the SED-based SFR, SFR$_\mathrm{SED}$ 
    (b: middle). (c) The right panel compares SFR$_\mathrm{UV,sed}$ 
    and SFR$_\mathrm{SED}$ with SFR$_\mathrm{UV,\beta}$, which is used 
    in the main body of this paper (Fig.~\ref{fig8}). 
    SFR$_\mathrm{UV,\beta}$ is derived from the UV luminosity with a dust 
    correction based on the UV slope. The symbols are the same as in
    Fig.~\ref{fig8}, but on the right panel, SFR$_\mathrm{UV,\beta}$ 
    and SFR$_\mathrm{SED}$ are represented by filled and open symbols, 
    respectively. Dashed, dot-dash, and solid lines represent the main 
    sequences of the star-forming galaxies at $z$=2.0--2.5 reported by 
    \citet{Whitaker:2014b}, the one at $z$=2.5 from the
    prescription by \citet{Tomczak:2016}, and the one at $z$=2.0--2.6 by
    \citet{Shivaei:2015}, respectively.
    The SFR calibrations used in \citet{Whitaker:2014b} and \citet{Tomczak:2016}
    is based on UV+IR luminosities.
    For the \citet{Shivaei:2015} main sequence, we use their SFRs estimated
    in the same manner for each diagram for consistency.
    One should note that these works present 
    the main sequence only for massive galaxies above $1\times10^{9}$ or 
    $1\times10^{10}$ \msun\ at $z\sim2.5$. Here we just extrapolate 
    their measured slopes to the lower mass galaxies.}
    \label{fig16}
\end{figure*}

Therefore, we carry out more comprehensive comparisons of the 
star-forming main sequence based on different recipes of dust 
corrections and SFR measurements other than \ha-based one, such as SFRs 
derived from UV luminosities with dust correction by the UV slope 
(\S2.5), the SED-fitting, or SFRs from full SED-fitting over a wide 
wavelength range. The latter method assumes the delayed exponentially 
declining star-formation histories and the other initial parameters are 
the same as \S2.5. The results are shown in Fig.~\ref{fig16}. Large 
error-bars associated to the SFR$_\mathrm{SED}$ of the LAEs are due to 
their faintness at all broad-band data as mentioned in \S2.5. However, 
most of the measured SFRs are consistent with each other 
(Fig.~\ref{fig16}c), and the median values agree with the main sequences 
presented for comparison which are obtained in the studies in the 
literature \citep{Whitaker:2014b,Shivaei:2015,Tomczak:2016}. 

\section{Selection of \lya\ Absorbers}

\begin{figure}
	\includegraphics[width=75mm]{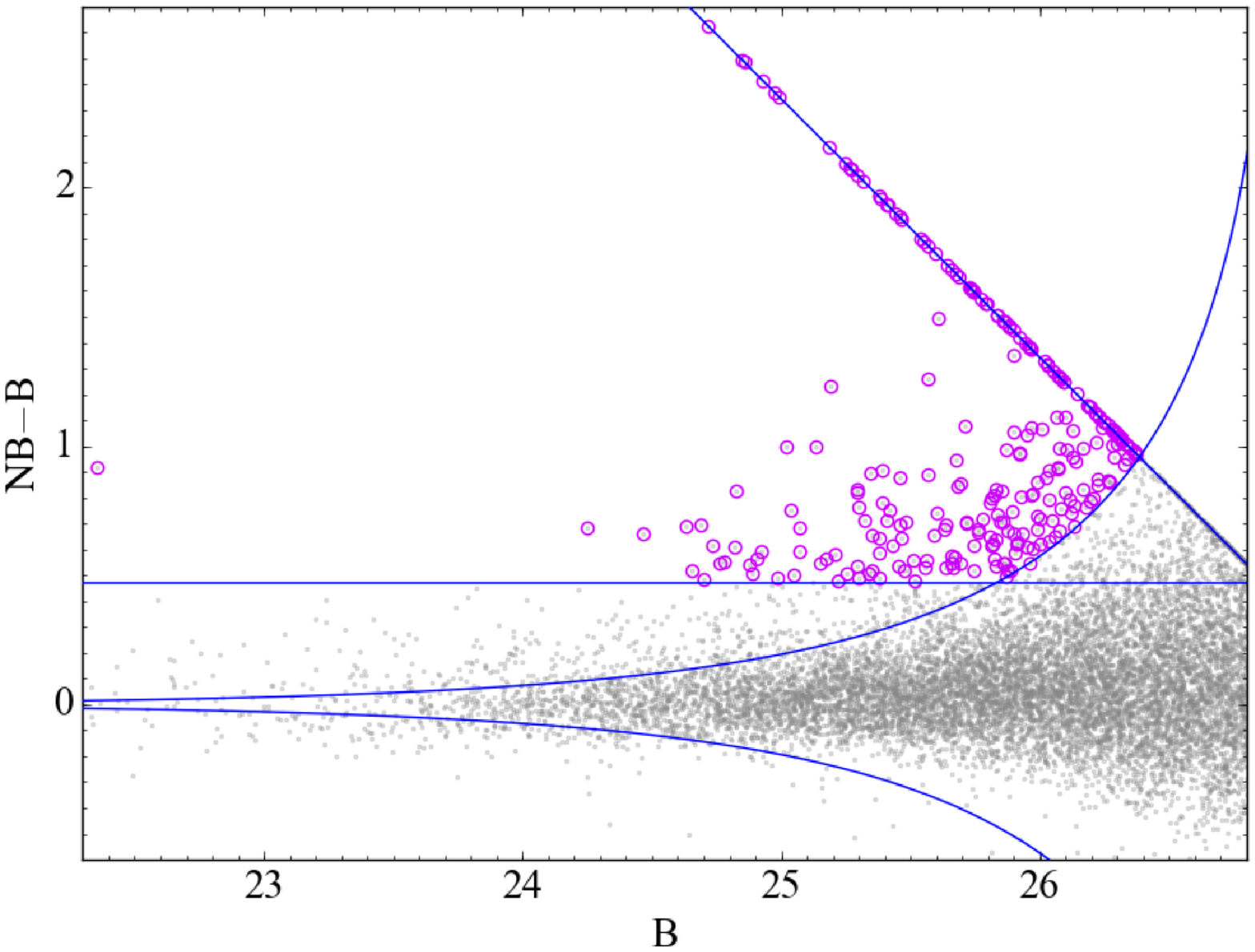}
    \caption{The colour-magnitude diagram of B vs. NB428$-$B. 
    The grey dots show all the B-detected galaxies in the UDS-WFC3
    field. The blue curves indicate the threshold of $\pm3\sigma$ 
    flux excess. The blue horizontal line indicates the 
    criterion used for the equivalent width cut corresponding
    to EW$_\mathrm{rest}=-9$ \AA. 
    The blue diagonal line indicates the 2$\sigma$ limit in B-band. The 
    purple open circles show the NB selected absorbers that satisfy our 
    selection criteria.}
    \label{fig17}
\end{figure}

This section provides supplementary information about the \lya\ 
absorption selection (\S4.4). Figure~\ref{fig17} shows the 
colour--magnitude diagram for narrowband absorbers where stars with 
F160W$<$25 mag classified by the \citet{Skelton:2014} catalogue have 
been removed. As described in the main text, we chose the absorbers 
showing more than $3\sigma$ deficits in their narrowband photometries 
with respect to those B-band magnitudes and have NB$-$B $>0.476$ mag 
and B $<27.89$ mag ($5\sigma$ limiting magnitude). 

In order to select LAAs at $z=2.53\pm0.03$, this work has employed 
photometric redshifts of the narrowband absorbers derived from the 
\citet{Momcheva:2016} photo-$z$ catalogue based on the 3D-HST data.
Also, we here investigate possible contaminations by using the single 
stellar population (SSP) model called GALAXEV \citep{Bruzual:2003}.
The metallicities are fixed to the solar abundance, and no dust
extinction is considered. The emission line components are not included. 

\begin{figure}
	\includegraphics[width=75mm]{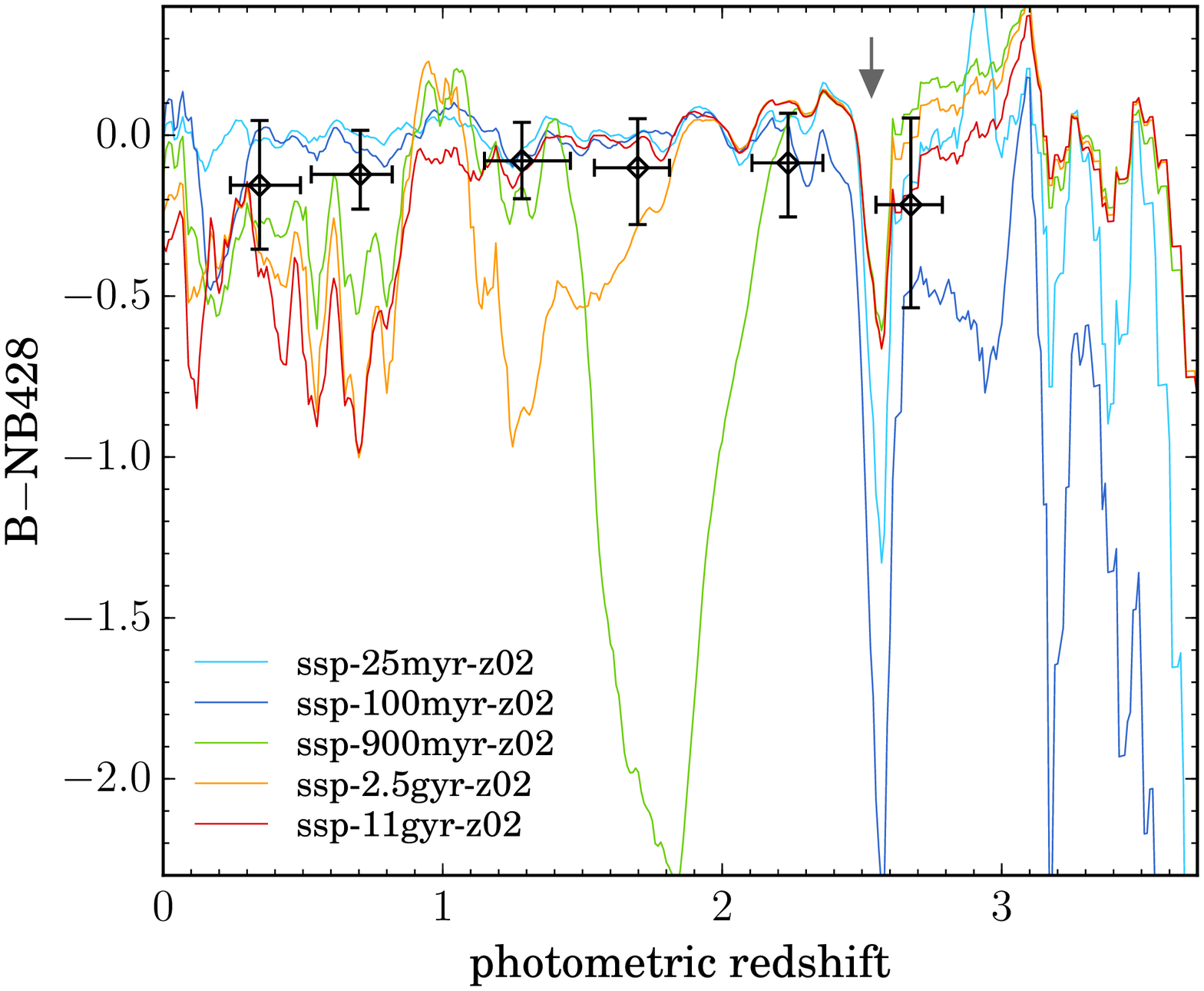}
    \caption{Colour term distributions of B$-$NB in the SSP models 
    with various ages, 0.025, 0.1, 0.9, 2.5, and 11 Gyrs, plotted as a function 
    of redshift. Those metallicities are fixed to the solar abundance. 
    The downward arrow indicates the \lya\ redshift ($z=2.53$) for the narrow-band.
    The black open circles and the errorbars show the median values and the
    $1\sigma$ scatters of the photo-$z$ sources, respectively.
    Those photometric redshifts are taken from the \citet{Momcheva:2016}
    catalogue ($z_\mathrm{best}$).}
    \label{fig18}
\end{figure}

The colour term distributions as a function of redshift are presented 
in Fig.~\ref{fig18}. Our colour term correction ($+0.1$ mag) in \S2.2 is
reasonable considering the typical colour-term values of 
B$-$NB$\sim$$-$0.1 at $z<2.5$.
Also, the simple models seem to be consistent with the
observed colour terms of the photo-$z$ determined galaxies 
\citep{Skelton:2014,Momcheva:2016} within the scatter. Slight 
downward offsets at $z>1.5$ would be due to strong dust attenuation 
in the FUV regime, since the central wavelength of the NB428 filter 
is slightly shorter by 170 \AA\ than that of the B-band. At $z>2.5$, 
any absorption by the intergalactic medium (IGM) makes the B$-$NB colour
bluer. Because of the difference in filter 
central wavelengths, a spectral break feature can also be captured
by the narrowband selection.
Especially, the UV break feature at around 2640 \AA\ by old stellar 
populations with age $\geq1$ Gyr is thought to be a main 
contamination as is actually seen in the photo-$z$ distribution 
of the narrow-band absorbers (Fig.~\ref{fig14}). 
Although there are strong dips between the redshift 1 and 2 in the 
SSP model with the age of 2.5 and 0.9 Gyrs, these features would 
be negligible for our selection because those flux densities at the 
narrowband wavelength are much lower by two order of magnitude than 
those in the rest-frame optical. 

We also see a dip in the colour-term distribution around $z=2.5$ due 
to the stellar \lya\ absorption. Galaxy spectra in the rest-frame FUV 
regime are mostly contributed from young massive stellar populations 
with the age younger than $\sim100$ Myr \citep{Bruzual:2003}.
The SSP models with the age of 25 and 100 Myrs show a significant 
narrow-band deficit due to a stellar \lya\ absorption at $z=2.5$,
suggesting a possibility that our narrow-band selection may pick up
this level of stellar absorption feature. In practice, however, 
such a young stellar population should also produce \lya\ line in emission,
and this absorption feature may be filled up and disappear 
as simulated by \citet{Pena:2013} (see the main text in \S4.4).

\bsp	
\label{lastpage}
\end{document}